\documentclass[12pt]{article}
\RequirePackage[authoryear]{natbib}

\RequirePackage{amsthm,amsmath,amsfonts,amssymb}
\RequirePackage[colorlinks,citecolor=blue,urlcolor=blue]{hyperref}
\RequirePackage{graphicx}

\usepackage{float}
\usepackage{enumitem}
\usepackage{url}
\usepackage{bbm}
\usepackage[nottoc,numbib]{tocbibind}
\usepackage{booktabs,caption,subcaption}
\usepackage[flushleft]{threeparttable}
\usepackage{multirow}

\usepackage[pdftex,dvipsnames]{xcolor}
\usepackage{comment}

\usepackage[margin=1in]{geometry}

\usepackage{macros}

\theoremstyle{definition}

\begin{document}


\title{\bf Multivariate and Online Transfer Learning with Uncertainty Quantification}
 \author{Jimmy Hickey, Jonathan P. Williams, Brian J. Reich, Emily C. Hector\hspace{.2cm}\\
  Department of Statistics, North Carolina State University}
  \date{}
 \maketitle
 
 \bigskip
\begin{abstract}
Untreated periodontitis causes inflammation within the supporting tissue of the teeth and can ultimately lead to tooth loss.
Modeling periodontal outcomes is beneficial as they are difficult and time consuming to measure, but disparities in representation between demographic groups must be considered.
There may not be enough participants to build group specific models and it can be ineffective, and even dangerous, to apply a model to participants in an underrepresented group if demographic differences were not considered during training.
We propose an extension to RECaST Bayesian transfer learning framework.
Our method jointly models multivariate outcomes, exhibiting significant improvement over the previous univariate RECaST method.
Further, we introduce an online approach to model sequential data sets.
Negative transfer is mitigated to ensure that the information shared from the other demographic groups does not negatively impact the modeling of the underrepresented participants.
The Bayesian framework naturally provides uncertainty quantification on predictions.
Especially important in medical applications, our method does not share data between domains.
We demonstrate the effectiveness of our method in both predictive performance and uncertainty quantification on simulated data and on a database of dental records from the HealthPartners Institute.
\end{abstract}

\noindent%

\section{Introduction}
\label{section: introduction}


The field of transfer learning has developed to solve machine learning problems where data or computational resources may be limited.
Transfer learning leverages information from \textit{source} domains where data are plentiful to aid in learning about related, resource scarce \textit{target} domains.
Importantly, while these domains are similar they are not necessarily the same.
If the domains are different, transfer learning methods may incur {\em negative transfer} where the inclusion of unrelated source information hinders the learning of the target task.

The Random Effect Calibration of Source to Target (RECaST) framework proposed in \cite{hickey2022transfer} is a Bayesian transfer learning method.
The RECaST framework represents the similarity between the source and target as a Cauchy-distributed random effect.
The posterior distribution of the random effect parameters is learned on the target data using a model built in the source domain.
This learned posterior distribution is used to construct predictions and posterior predictive credible sets for unseen target data -- providing both prediction and uncertainty quantification without necessitating the access to the source data.

The HealthPartners Institute at Minneapolis, Minnesota has collected a longitudinal, observational data set of their participants through routine dental checkups as described in \cite{guan2020bayesian}.
The data consider demographic and dental features along with two outcomes that are indicators of periodontal disease.
Certain racial groups are much less represented than others in the data.
Ignoring this imbalance and using all of the data jointly to build a single model could result in poor predictive predictive performance and uncertainty quantification on underrepresented populations.
We address this disparity by sharing information across related groups with multivariate online RECaST.
The proposed methods enable us to model both outcomes jointly and to create better models for underrepresented racial groups.
While transfer learning methods to model online or multivariate data exist, no others provide uncertainty quantification on the predictions.

We extend the RECaST framework in two ways.
First, we model multivariate outcomes whereas the original framework was limited to univariate outcomes.
We provide two different multivariate random effect distributions to model the similarity between source and target.
One uses a multivariate Cauchy distribution, a natural extension from the univariate Cauchy.
The other incorporates the univariate Cauchy marginal distributions of the original RECaST framework with a multivariate Gaussian copula.
Second, we propose an online method to share information between sequential target data sets while mitigating negative transfer.
As we demonstrate in the data analysis, this online analysis can also be applied to data that are collected concurrently but modeled sequentially.

The newly proposed methods retain many of the advantages of the original RECaST framework.
Data privacy is maintained between all data sets.
Only the source model, not the source data, is required when fitting the posterior distribution for the target data.
When considering multiple target data sets, only the learned posterior distribution from previous targets are shared and not the data themselves.
This property is especially important when working with sensitive data such as medical information.
Additionally, the framework remains agnostic to the specification of the source model, such that it can correspond to a parametric generalized linear model or a highly flexible neural network.

The RECaST methods are evaluated on both predictive performance and posterior predictive coverage.
We consider synthetic simulation studies that vary sample size, outcome correlation, and similarity between source and target data sets.
Changes in these parameters represent different challenges of real data sets and enables us to isolate the effects that each parameter has on predictive performance and uncertainty quantification. 
Negative transfer is examined throughout the studies.

The remainder of this paper is organized as follows.
In Section \ref{section: related work} we discuss related work in transfer learning.
We provide a summary of the original RECaST framework that this work builds upon in Section \ref{section: recap}.
We extend this framework in Section \ref{section: multivariate} to model multivariate outcomes, providing two natural modeling choices.
In Section \ref{section: online}, we develop an online extension to handle sequential target data sets.
We test these methods in Section \ref{section: simulation study} through extensive simulation studies.
Section \ref{section: dental data} analyzes the periodontal disease outcomes.
Section \ref{section: concluding remarks} concludes.
Further mathematical and computational details can be found in the Appendix.

\section{Related Work}
\label{section: related work}

\cite{pan2009}, \cite{lu2015}, \cite{weiss2016}, \cite{dube2020}, \cite{zhuang2021} are general survey papers on transfer learning.
See \cite{suder2023bayesian} for a specialized survey of Bayesian transfer learning methods.

Using transfer learning for underrepresented demographic groups can provide significant improvement when compared to building models where demographic differences are ignored.
TransRF is a random forest based approach to predict breast cancer using European women as the source population and patients with African and South Asian ancestry as the target population \citep{gu2022transfer}.
In \cite{HONG2024104532}, a LASSO based federated transfer learning method is proposed to address demographic and outcome class imbalance for stroke risk prediction.
In \cite{li2023targeting}, ideas from sparse high-dimensional regression are used to construct polygenic risk scores for Type II diabetes for patients in underrepresented racial groups using genome-wide association studies.
The STRIFLE method builds upon this by including nuisance parameters which provide robustness to negative transfer \citep{cai2024semi}.

Multivariate outcomes are common in transfer learning, especially for computer vision tasks. 
A deep convolutional neural network built to classify images on a fixed set of outcome classes is generalized to predict outcomes in other domains in \cite{donahue2014decaf}.
In \cite{NEURIPS2021_a0d3973a}, a method is proposed to handle posterior sampling for a large number outcomes which incurs expensive sampling because of a large covariance matrix in .
In \cite{Singh_2023_ICCV}, the standard approach of pre-training a neural network in the source domain and fine-tuning in the target domain is expanded to include an additional pre-pre-training step to improve foundational models.
Sequential improvement through online image data sets is considered in \cite{Yu_2024_CVPR}.
These methods focus exclusively on prediction while multivariate RECaST also provides valuable uncertainty quantification.

Multitask learning uses information from a source domain to improve performance on multiple target domains \citep{caruana1997multitask}.
In \cite{bakker2003task}, a Bayesian neural network method is proposed that fits the output network weights with task-specific data, but uses data from all domains to learn shared parameters for the rest of the network.
A multitask Bayesian optimization for Gaussian process models that focuses on efficiently handling the exploitation-exploration trade-off of parameter optimization to jointly minimize error over all tasks is developed in \cite{NIPS2013_f33ba15e}.
These methods generally optimize over all of the target tasks jointly.
In contrast, the proposed online RECaST method learns a posterior distribution specific to each target task.

Online and online-batch transfer learning methods focus on updating model parameters as data sets arrive sequentially.
This is in contrast to batch learning which requires all of the data to be present before training can occur.
\cite{opper1999bayesian} develop an early approach to updating a Bayesian posterior distribution as new data points arrive, boasting asymptotic efficiency  with Gaussian families.
An online Bayesian method that combines models from multiple data sets is proposed by \cite{chen2001approach}.
\cite{wu2017online} perform online transfer learning by weighing models built on multiple source domains in an ensemble in combination with a target domain classifier to improve prediction in the target domain.
In \cite{wu2023bayesian}, performance bounds are calculated based on source and target sample sizes to choose informative Bayesian priors when the target domain data arrive in one batch, online, or in sequential batches.
In \cite{pmlr-v202-patel23a}, a federated, online approach is proposed to a sequential decision making framework where communication between sites is limited.
A weighting scheme is developed for online data across multiple sites to control the trade-off between training time and bias in \cite{pmlr-v206-marfoq23a}.

Some Bayesian transfer learning methods focus on learning an informative prior to improve prediction on a target task.
\cite{raina2006constructing} propose a method to learn a covariance matrix from source text classification tasks which is used as a prior on the covariance matrix in the target domain.
In \cite{kapoor2021variational}, a continual learning approach to Bayesian inference is developed that updates a posterior distribution as new tasks arrive using a Gaussian process.
A tuning parameter that adapts the posterior learned on a source task to be used as the prior on a target task is introduced in \cite{shwartz2022pre}.
This is similar to the proposed online RECaST with some key differences: online RECaST uses posterior information from previous {\em target} data sets in the prior distribution and introduces a weighting parameter that is {\em learned} rather than tuned. 
In \cite{abba2022penalizedcomplexitypriordeep}, a penalized complexity prior based on the Kullback–Leibler divergence between the source and target models is developed.
A shrinkage estimator is used in the prior when the difference between the source and target tasks is sparse in \cite{abba2024bayesianshrinkageestimatortransfer}.
The online RECaST method that we propose uses a prior informed by previous target data sets while mitigating negative transfer.


\section{Summary of the RECaST Framework}
\label{section: recap}

We begin with a summary of the RECaST framework from \cite{hickey2022transfer}.
RECaST is a Bayesian transfer learning method for source and target data sets that share the same outcome and feature spaces but may have differences in feature-to-outcome mappings.
It is scalable, requiring estimation of only 2-3 parameters, which is especially important when there are minimal target data.
RECaST maintains data privacy, requiring only the fitted source model to be shared and not the source data.
The estimation of a Bayesian posterior distribution naturally begets uncertainty quantification through posterior predictive credible intervals.

Take $Y_{S}$ to be an observation from the source data set with corresponding features $\bx_{S}$ and $Y_{T}$ to be an observation from the target data set with features $\bx_{T}.$
Denote the forward data-generating mechanism of the feature-to-outcome mappings $P_{S}(y_{S} \mid \bx_{S})$ and $P_{T}(y_{T} \mid \bx_{T})$, respectively, by
\begin{align*}
Y_{S} = h \Big\{ f(\btheta_{S}, \bx_{S}), U_{S} \Big\} \hspace{0.25cm} \text{ and } \hspace{0.25cm}
Y_{T} = h \Big\{ g(\btheta_{T}, \bx_{T}) , U_{T} \Big\},
\end{align*}
where $f(\btheta_{S}, \bx_{S})$ and $g(\btheta_{T}, \bx_{T})$ are the \textit{structural components} that relate an observation's features $\bx$ to the data generating parameters $\btheta$.
The $h$ function relates these structural components to $U_{S}$ and $U_{T}$, which are independent auxiliary random variables.
For example,
\begin{align*}
f(\btheta_{S}, \bx_{S}) & = \bx_{S}^{\top}\btheta_{S}, \\
h(\bx_{S}^{\top}\btheta_{S}, U_{S}) & = f(\btheta_{S}, \bx_{S}) + U_{S}, \ \ \text{and} \\
U_{S} & \sim \mathcal{N}(0,1),
\end{align*}
then $Y_{S} \sim \mathcal{N}(\bx_{S}^{\top}\btheta_{S}, 1)$.
With a small target sample size it may be infeasible to estimate $g(\btheta_{T},\bx_{T})$ making transfer learning necessary to model the target task.

The RECaST framework defines $\beta := g(\btheta_{T}, \bx_{T}) / f(\btheta_{S}, \bx_{T})$ which intuitively represents the \textit{similarity} between the source and target data generating mechanisms.
If the source and target are generated in a similar way, then $\beta \approx 1$.
With this new $\beta$ term, the target generating mechanism can be expressed as
\begin{align*}
Y_{T,i} & = h \Big\{ g(\btheta_{T}, \bx_{T,i}), U_{T,i} \Big\} \\
   & =  h \Big\{ \frac{f(\btheta_{S}, \bx_{T,i})}{f(\btheta_{S}, \bx_{T,i})} \cdot g(\btheta_{T}, \bx_{T,i}), U_{T,i} \Big\} \\
  & =  h \Big\{ \beta_{i} \cdot f(\btheta_{S}, \bx_{T,i}), U_{T,i} \Big\}
\end{align*}
for $i \in \{ 1, \dots , n_{T} \}$, where $Y_{T,1} , \dots , Y_{T, n_{T}}$ is an independent sample of $n_{T}$ target outcomes with associated features $\bx_{T,1} , \dots , \bx_{T, n_{T}}$ and $\beta_{i} = g(\btheta_{T}, \bx_{T,i}) / f(\btheta_{S}, \bx_{T,i})$.
Notice that now the model for $Y_{T,i}$ depends on $f(\btheta_{S},\bx_{T,i})$, which is assumed to have a reliable estimate from the source, and $\beta_{i}$.
The reliance on estimating $g(\btheta_{T}, \bx_{T,i})$ has been replaced with $\beta_{i}$ which is modeled as a random effect.

Lemma 1 of \cite{hickey2022transfer} states in the canonical case where $\bx_{T,1}, \dots , \bx_{T,n_{T}} \overset{\text{iid}}{\sim} \mathcal{N}_{p}(\pmb{0}, \bI_{p})$, and $g(\btheta, \bx) = f(\btheta, \bx) = \bx^{\top} \btheta$ that $\beta_{i} \sim \text{Cauchy}(\delta, \gamma)$ with
\begin{align}
    \delta  = \frac{\btheta_{T}^{\top} \btheta_{S}}{ \lVert \btheta_{S} \rVert ^{2}} \hspace{0.5cm} \text{ and } \hspace{0.5cm} \gamma  = \frac{1}{\lVert \btheta_{S} \rVert^{2}} \sqrt{ \lVert \btheta_{S} \rVert ^{2} \lVert \btheta_{T} \rVert ^{2} - (\btheta_{T}^{\top} \btheta_{S})^{2} }.
    \label{equation: recast canonical}
\end{align}
In this case, the distribution of $\beta_{i}$ exactly follows a Cauchy distribution.
In fact, the heavy tails of the Cauchy distribution also make it practically useful for situations where those canonical assumptions are violated.
Since $\beta_{i}$ characterizes the similarity between the source and the target domains, these heavy tails allow the random effect to capture large disparities between the source and target.
RECaST models the $\beta_{i}$ terms as random effects following a $\text{Cauchy}(\delta, \gamma)$ distribution with shared parameters $\delta$ and $\gamma$.
The joint posterior distribution of $(\delta, \gamma)$ is learned using the target data and the pre-fitted source model.
RECaST parameters $\delta$ and $\gamma$ can be learned using standard Markov chain Monte Carlo methods even outside of the canonical case.
The Bayesian posterior predictive distribution can be sampled to generate empirical credible intervals for uncertainty quantification.


\section{Multivariate RECaST}
\label{section: multivariate}

\subsection{Data Generating Mechanism}
\label{section: data generating mechanism}

We now extend the RECaST framework to model multivariate outcomes.
Take  $\by = [y_{1}, \dots , y_{m}]^{\top}$ to be a vector of $m$ outcomes, $\bx = [x_{1}, \dots , x_{p}]$ to be the corresponding vector of $p$ features, and $\bTheta$ to be the generating parameters.
The forward data-generating mechanisms $P_{S}(\by_{S} \mid \bx_{S})$ and $P_{T}(\by_{T} \mid \bx_{T})$ can be represented, respectively, by
\begin{align*}
    \bY_{S} = h \Big\{ f(\bTheta_{S} , \bx_{S}), \bU_{S} \Big\} \hspace{0.5cm} \text{ and } \hspace{0.5cm}
    \bY_{T} = h \Big\{ g(\bTheta_{T} , \bx_{T}), \bU_{T} \Big\},
\end{align*}
where $\bU_{S}$ and $\bU_{T}$ are independent auxiliary random variables with dimension $m \times m$.

As before, the transfer learning question arises when there is not enough target data to reliably estimate $g(\bTheta_{T}, \bx_{T})$.
To capture the similarity between the source and target domains we define 
$\beta_{i,j} := g(\bTheta_{T,j}, \bx_{T,i}) / f(\bTheta_{S,j}, \bx_{T,i})$ for $i = 1, \dots ,n_{T}$ and $j = 1, \dots , m$.
In words, for each target observation $i$, there is a $\beta_{i, j}$ ratio for each outcome that models the relation between the source and the target outcome.
Take $\bbeta_{i} = [\beta_{i, 1}, \dots , \beta_{i, m}]^{\top}$  to be the vector of ratios and $\text{diag}(\bbeta)$ to be the matrix with elements of $\bbeta$ on the diagonal and $0$'s elsewhere.
With this, we now express the target generating mechanism as%
\begin{align*}
    \bY_{T,i} & = h\Big\{ g(\bTheta_{T}, \bx_{T,i}), \bU_{T,i} \Big\} \\
        & = h \Big\{ \text{diag}(\bbeta_{i}) f(\bTheta_{S}, \bx_{T,i}), \bU_{T,i}\Big\}.
\end{align*}

Thus, $\bY_{T,i}$ is no longer expressed as a function of $g(\bTheta_{T}, \bx_{T,i})$, which cannot be reliably estimated.
The function $g(\bTheta_{T}, \bx_{T,i})$ is replaced by $f(\bTheta_{S}, \bx_{T,i})$, which we assume can be reliably estimated from the source domain, and $\bbeta_{i}$, the random effect for individual $i$, $i = 1, \dots n_{T}$.
In addition to removing the dependence on $g(\bTheta_{T}, \bx_{T,i})$, considering 
The relationship between the outcomes in the target domain will be estimated through the parameters of $\bU_{T,i}$.

Considering ratios of different outcomes, such as $g(\bTheta_{T,j}, \bx_{T,i}) / f(\bTheta_{S,k}, \bx_{T,i})$ for $j \neq k$, would result in a non-diagonal matrix multiplied by $f(\bTheta_{S}, \bx_{T,i})$ and thus would not remove the dependence on $g(\bTheta_{T}, \bx_{T,i})$.
Additionally, considering only ratios makes the $\beta_{i,j}$ terms interpretable: if the source and target are generated from the same distribution, then $\bbeta_{i} = \boldsymbol{1}_{m}$, a vector of 1's.
This mirrors $\beta = 1$ in univariate RECaST described in Section \ref{section: recap}.
It is unclear what the expected behavior would be if a ratio of different outcomes were considered.

Unlike univariate RECaST, the multivariate random effect $\bbeta_{i}$ does not always follow a multivariate Cauchy distribution, even in the canonical case.
A vector of random variables follows a multivariate Cauchy distribution if and only if every linear combination of the components follows a univariate Cauchy distribution.
Because of the covariance between the elements of $\bbeta_{i}$, this is not always the case \citep{pillai2016ratios, pillai2016unexpected}.
Thus, we provide two choices of random effect distributions for $\bbeta_{i}$.
The first is a multivariate Cauchy distribution.
While this may not always be the exact distribution of $\bbeta_{i}$, it is a natural extension of the univariate model.
In practice, the heavy tails are, again, beneficial for capturing the relationship between source and target even if the underlying distributions differ.
Second, we propose a copula-based approach that leverages the fact that, in the canonical case, the marginal distributions of the elements of $\bbeta_{i}$ are known to be univariate Cauchy.
A multivariate Gaussian copula is used to model the dependence structure.
In practice, these models can be used regardless of the form of the source model $f(\bTheta_{S}, \bx)$.

\subsection{Multivariate Cauchy}
Univariate RECaST models the scalar $\beta$ term with a Cauchy distribution.
This follows from $\beta$ representing the ratio of two normally distributed random variables in the canonical case.
The natural extension for a multi-dimensional outcome is to model $\bbeta$ with an $m$-dimensional multivariate Cauchy distribution: $\bbeta_{i} \sim \text{Cauchy}_{m}(\bdelta, \bGamma)$. Here $\bdelta \in \mathbb{R}^{m \times 1}$ is the location vector and $\bGamma$ is an $m \times m$ positive definite scale matrix, $i \in \{1, \dots , n_{T} \}$.
If the source and target come from similar distributions we expect the elements of $\bdelta$ to be close to 1.
The $\bGamma$ matrix captures the dependence between the elements of $\bbeta_{i}$.
For continuous responses $\by \in \mathbb{R}^{m \times 1}$, a natural choice for the $h$ function is the Gaussian innovation function,
\begin{align*}
    \by_{T,i} & = \text{diag}(\bbeta_{i}) f(\widehat{\bTheta}_{S}, \bx_{T,i})  + \bU_{T,i},
\end{align*}
where $\bU_{T,i} \sim \mathcal{N}_{m}(0, \bSigma)$, $\bSigma$ is an unknown $m \times m$ covariance matrix, and $\widehat{\bTheta}_{S}$ are parameter estimates from the source model.
Denoting $\pi(\bdelta, \bGamma, \bSigma)$ as a prior density on $(\bdelta, \bGamma , \bSigma)$, the posterior distribution of $(\bdelta, \bGamma , \bSigma)$ is
\begin{align*}
p&(\bdelta, \bGamma, \bSigma \mid \by_{T,1} , \dots , \by_{T,n_{T}} , \widehat{\bTheta}_{S}) \\
%
%
    & \propto \pi( \bdelta, \bGamma, \bSigma ) \cdot \prod_{i=1}^{n_{T}} \int_{\mathbb{R}^{m}} \mathcal{N}_{m} \{ \by_{T,i} \mid \text{diag}(\bbeta_{i}) f(\widehat{\bTheta}_{S}, \bx_{T,i}), \ \bSigma \} \cdot \text{Cauchy}_{m}(\bbeta_{i} \mid \bdelta, \bGamma) \ d \bbeta_{i}.
    \numberthis \label{equation: mv cauchy posterior}
\end{align*}
Since $\bdelta$ is a real valued location vector, $\bGamma$ is a scale matrix, and $\bSigma$ is a covariance matrix, a canonical choice of prior distribution is
\begin{align*}
    \pi(\bdelta, \bGamma, \bSigma) = \mathcal{N}_{m}(\bdelta \mid \pmb{1}_{m}, \bSigma_{\bdelta}) \cdot \text{IW}_{m}( \bGamma \mid \bPsi_{\bGamma}, \nu_{\bGamma}) \cdot \text{IW}_{m}(\bSigma \mid \bPsi_{\bSigma}, \nu_{\bSigma}),
\end{align*}
where IW is the inverse-Wishart distribution.
The hyperparameters for $\bSigma$ can be chosen based on prior information known about the covariance of the outcomes in the target domain.
The hyperparameters for $\bdelta$ and $\bGamma$ can be chosen based on the relationship between the source and target domains; if they are known to be similar then a small covariance for $\bdelta$ and a mean near the identity for $\bGamma$ will result in a prior that favors $\bbeta$ being near $\boldsymbol{1}_{m}$.
In practice, we choose hyperparameters that are diffuse to demonstrate that RECaST can still perform well on unseen target test data without prior information.

The posterior distribution is estimated using the random walk Metropolis-Hastings algorithm using the \texttt{nimble} package for \texttt{R} \citep{nimble2016}.
The total number of iterations $n_{\text{iterations}}$ and number of burn-in steps can be chosen based on available computational resources at the target site or stopped early based on convergence.
The computational complexity is $\mathcal{O}(n_{T} \cdot n_{\text{iterations}})$.
This alleviates concerns about scalability since $n_{T}$ is assumed to be small for transfer learning problems.
For this choice of random effect distribution, a Gibbs sampler could also be used; the details for this, including the full conditional distributions, are outlined in Appendix \ref{section: gibbs}.
Once the posterior distribution is estimated, the posterior predictive distribution is constructed as follows.

Take $\widetilde{ \by }_{T}$ to be the outcome vector of a newly observed target observation with corresponding features $\widetilde{ \bx }_{T}$.
The posterior predictive distribution of $\widetilde{ \by }_{T}$ is the marginal distribution of 
\begin{align*}
\pi & (\widetilde{ \by }_{T}, \widetilde{ \bbeta }, \bdelta, \bGamma, \bSigma \mid \by_{T,1}, \dots \by_{T, n_{T}}, \widehat{ \bTheta }_{S}) \\
%
    %
    & = \mathcal{N}_{m} \big\{\widetilde{ \by }_{T} \mid \text{diag}(\widetilde{ \bbeta }) f(\widehat{ \bTheta }_{S}, \widetilde{ \bx }_{T}), \bSigma \big\} \cdot \text{Cauchy}_{m}(\widetilde{ \bB } \mid \bdelta, \bGamma) \cdot p(\bdelta, \bGamma, \bSigma \mid \by_{T,1}, \dots \by_{T, n_{T}}).
\end{align*}
Algorithm \ref{algorithm: multivariate cauchy RECaST prediction} describes a procedure for drawing posterior predictive samples from this distribution.
%
This procedure gives a sample of outcome vectors for the newly observed features $\widetilde{ \bx }$ of size $n_{\text{post}} \cdot n_{\bbeta} \cdot n_{\bY}$.
Correspondingly, the computational complexity is $\mathcal{O}(n_{\text{post}} \cdot n_{\bbeta} \cdot n_{\bY})$; the sampling parameters can be adjusted based on available computing resources.
Given a nominal coverage level $\alpha$, we create elliptical posterior predictive credible sets from these samples. 
For each sampled outcome, we compute the Mahalanobis distance to the mean of the samples; denote this set of Mahalanobis distances as $\pmb{\text{MH}}$.
The outcome is covered if the Mahalanobis distance between the true value $\widetilde{ \by }_{T}$ and the mean of the sampled outcomes is within the inner $1 - \alpha$ percentile interval of $\pmb{\text{MH}}$.

\begin{algorithm}[H]
\small
 \caption{\footnotesize Multivariate Cauchy RECaST Posterior Predictive Sampling}\label{algorithm: multivariate cauchy RECaST prediction}
 \textbf{Input:} $\widetilde{\bx}_{T}$, an estimated posterior $p(\bdelta, \bGamma, \bSigma \mid \by_{T,1}, \dots \by_{T, n_{T}})$, and sample sizes $n_{\text{post}}$, $n_{\bbeta}$, and $n_{Y}$\\
\textbf{Output:} A sample of values from $\pi\big(\widetilde{\by}_{T} \mid \by_{T,1}, \dots, \by_{T,n_{T}}, \widehat{\bTheta}_{S} \big)$
 \begin{algorithmic}

 \For{$i \gets 1$ to $n_{\text{post}}$}
     \State $\bdelta, \bGamma, \bSigma \gets \text{random} \big\{ p\big( \bdelta, \bGamma, \bSigma \mid \by_{T,1}, \dots, \by_{T,n_{T}}, \widehat{\btheta}_{S}\big)  \big\}$
     \For{$j \gets 1$ to $n_{\bbeta}$}
         \State $\widetilde{ \bbeta } \gets \text{random}\big\{ \text{Cauchy}_{m}(\bdelta, \bGamma)  \big\}$
         \For{$k \gets 1$ to $n_{\bY}$}
             \State $\widetilde{\bY}_{T} \gets \text{random}\big[  \mathcal{N}_{m}\big\{\widetilde{ \by }_{T} \mid \text{diag}(\widetilde{ \bbeta }) f(\widehat{ \bTheta }_{S}, \widetilde{ \bx }_{T}), \bSigma \big\}\big]$
         \EndFor
     \EndFor
 \EndFor
 \end{algorithmic}
\end{algorithm}

\subsection{Multivariate Normal Copula with Cauchy Marginals}
Copulas are a modeling approach that construct a joint distribution from marginal distributions \citep{nelsen2006introduction}.
From univariate RECaST we know that the ratios $\beta_{j} = (\bTheta_{T,j} \bx_{T}) / (\bTheta_{S,j} \bx_{T}) \sim \text{Cauchy}(\delta_{j}, \gamma_{j})$ in the canonical case where $j = 1, \dots, m$ indexes the outcomes. 
That is, the marginal distributions of the elements of $\bbeta$ are known to be univariate Cauchy with their own center and scale parameters $\delta_{j}$ and $\gamma_{j}$, respectively. 
As with univariate RECaST, this canonical linear form is used to motivate the choice of parametric family but is not required to apply RECaST.
Define $u_{j} := F_{\beta_{j}}(\beta_{j})$ where $F_{\beta_{j}}(\cdot)$ is the cumulative distribution function of the Cauchy distribution associated with $\beta_{j}$.
By the probability integral transform, $u_{i,j} \sim U(0,1)$.
We use a Gaussian copula centered at $\boldsymbol{0}_{m}$ with correlation matrix $\bR$ to capture the dependence between the $\beta_{j}$'s.
The joint distribution function of the $\beta_{j}$'s is given by
\begin{align*}
    \pi & (\beta_{1}, \dots , \beta_{m} \mid \bR, \delta_{1}, \dots ,\delta_{m}, \gamma_{1} , \dots , \gamma_{m} ) \\
    & =  \prod_{j=1}^{m}  f_{\beta{j}}(\beta_{j})  \cdot c(u_{1}, \dots , u_{m} \mid \bR, \delta_{1}, \dots , \delta_{m}, \gamma_{1} , \dots , \gamma_{m}) \\
        & = \prod_{j = 1}^{m} \Big[ f_{\beta{j}}\{F_{\beta_{j}}^{-1}(u_{j})\} \Big] \cdot \mathcal{N}_{m}\Big(
         \begin{bmatrix}
        \Phi^{-1}\{ F_{\beta_{1}}(\beta_{1}) \} & 
        \dots &
        \Phi^{-1} \{ F_{\beta_{m}}(\beta_{m}) \} \mid \boldsymbol{0}_{m}, \bR
    \end{bmatrix}
        \Big),
\end{align*}
where $\Phi^{-1}(\cdot)$ is the inverse of the standard Gaussian cumulative distribution function.
The unknown parameters are the center parameters for the Cauchy marginal distributions $\delta_{1}, \dots , \delta_{m}$, the scale parameters for the Cauchy marginal distributions $\gamma_{1} , \dots , \gamma_{m}$, the correlation matrix for the multivariate Gaussian copula over the random effects $\bR$, and the covariance matrix over the target outcomes $\bSigma$.
The posterior distribution for these unknown parameters is
\begin{align*}
    p & (\delta_{1}, \dots , \delta_{m} , \gamma_{1}, \dots , \gamma_{m}, \bR, \bSigma \mid \by_{1}, \dots , \by_{n_{T}}, \widehat{\bTheta}_{S}) \\
    & \propto \pi(\delta_{1}, \dots , \delta_{m} , \gamma_{1}, \dots , \gamma_{m}, \bR, \bSigma) \cdot  \\
    & \hspace{1cm} \prod_{i=1}^{n_{T}} \Big\{ \int_{\mathbb{R}} \cdots \int_{\mathbb{R}}  \pi  (\by_{i} \mid \beta_{i,1}, \dots ,\beta_{i, m}, \bSigma  , \widehat{\bTheta}_{S}) \cdot\\
    & \hspace{2cm} \pi(\beta_{i,1}, \dots ,\beta_{i, m} \mid \delta_{1}, \dots , \delta_{m} , \gamma_{1}, \dots , \gamma_{m}, \bR)     
    \ d\beta_{i,1} \dots \ d\beta_{i, m}\Big\}.
    \numberthis \label{equation: copula posterior}
\end{align*}
As before, the posterior is estimated using the Metropolis-Hastings algorithm.
The integrals of this posterior distribution can also be calculated with respect to the $u_{i,j}$; this may have computational advantages since the integrals are on the bounded interval $[0,1]$.
The conversion of these integrals is provided in Appendix \ref{section: copula integrals}.
A canonical choice of prior for the unknown parameters is
\begin{align*}
    \pi & (\delta_{1}, \dots , \delta_{m} , \gamma_{1}, \dots , \gamma_{m}, \bR, \bSigma) \\
     = \text{IW}_{m} & ( \bR \mid \bPsi_{\bR}, \nu_{\bR}) \cdot
    \text{IW}_{m}(\bSigma \mid \bPsi_{\bSigma}, \nu_{\bSigma}) \cdot 
     \prod_{j = 1}^{m} \mathcal{N}(\delta_{j} \mid \mu_{\delta_{j}}, \sigma^{2}_{\delta_{j}}) \cdot \text{IG}(\gamma_{j} \mid a_{\gamma_{j}}, b_{\gamma_{j}}),
\end{align*}
where $\text{IG}$ is the inverse-gamma distribution.

The posterior predictive distribution for a new target observation $\widetilde{ \by }_{T}$ with associated features $\widetilde{ \bx }_{T}$ is the marginal distribution of
\begin{align*}
    \pi & ( \widetilde{ \by }_{T}, \widetilde{ \beta }_{1}, \dots , \widetilde{ \beta }_{m} , \delta_{1}, \dots , \delta_{m} , \gamma_{1} , \dots , \gamma_{m}, \bR , \bSigma \mid \by_{1} , \dots , \by_{n_{T}} , \widehat{ \bTheta }_{S} ) \\
    %
    & = \pi( \widetilde{ \by }_{T} \mid \widetilde{ \beta }_{1}, \dots , \widetilde{ \beta }_{m} , \bSigma ) 
    \cdot 
    \pi( \widetilde{ \beta }_{1}, \dots , \widetilde{ \beta }_{m}    \mid \delta_{1}, \dots , \delta_{m} , \gamma_{1} , \dots , \gamma_{m}, \bR ) \cdot \\
    & \hspace{0.5cm} \pi(\delta_{1}, \dots , \delta_{m} , \gamma_{1} , \dots , \gamma_{m}, \bR , \bSigma  \mid \by_{1} , \dots , \by_{n_{T}} , \widehat{ \bTheta }_{S}).
\end{align*}
Algorithm \ref{algorithm: copula RECaST prediction} describes how to sample outcomes from this posterior distribution.
These samples can be used to create posterior predictive credible intervals and calculate empirical coverages as discussed in the previous section.
\begin{algorithm}[H]
\small
 \caption{\footnotesize Multivariate Copula RECaST Posterior Predictive Sampling}\label{algorithm: copula RECaST prediction}
 \textbf{Input:} $\widetilde{\bx}_{T}$, an estimated posterior $p(\delta_{1}, \dots , \delta_{m} , \gamma_{1}, \dots , \gamma_{m}, \bR, \bSigma \mid \by_{1}, \dots , \by_{n_{T}}, \widehat{\bTheta}_{S})$, and sample sizes $n_{\text{post}}$, $n_{\bbeta}$, and $n_{Y}$\\
\textbf{Output:} A sample of values from $\pi\big(\widetilde{\by}_{T} \mid \by_{T,1}, \dots, \by_{T,n_{T}}, \widehat{\bTheta}_{S} \big)$
 \begin{algorithmic}

 \For{$i \gets 1$ to $n_{\text{post}}$}
     \State $\delta_{1}, \dots , \delta_{m} , \gamma_{1}, \dots , \gamma_{m}, \bR, \bSigma \gets \text{random} \big\{ p(\delta_{1}, \dots , \delta_{m} , \gamma_{1}, \dots , \gamma_{m}, \bR, \bSigma \mid \by_{1}, \dots , \by_{n_{T}}, \widehat{\bTheta}_{S}) \big\}$
     \For{$j \gets 1$ to $n_{\bbeta}$}
         \State $\widetilde{ \beta }_{1} \gets \text{random}\big\{ \text{Cauchy}(\delta_{1}, \gamma_{1})  \big\}$
             
                \State\hspace{\algorithmicindent}
                $\vdots$
         
         \State $\widetilde{ \beta }_{m} \gets \text{random}\big\{ \text{Cauchy}(\delta_{m}, \gamma_{m})  \big\}$
         \For{$k \gets 1$ to $n_{\bY}$}
             \State $\widetilde{\bY}_{T} \gets \text{random}\big[  \mathcal{N}_{m}\big\{\widetilde{ \by }_{T} \mid \widetilde{ \bB } f(\widehat{ \bTheta }_{S}, \widetilde{ \bx }_{T}), \bSigma \big\}\big]$
         \EndFor
     \EndFor
 \EndFor
 \end{algorithmic}
\end{algorithm}


\section{Online RECaST}
\label{section: online}

We further build upon the RECaST framework to model multiple sequential target data sets.
We address the problem of online transfer learning, with multiple target domains, each of which may be lacking in sample size or computational resources.
As with previous RECaST methods, we operate under the assumption that the data are private and cannot be shared across sites, even between targets.
We propose a sequential approach that includes a RECaST model built in one target domain to inform a new target domain.

We motivate the problem with an example.
A hospital has data from the past 10 years to build a model that reliably diagnoses whether a patient has the flu -- this will act as the source.
Now the hospital wants to use that model on incoming patients for this year's flu season -- this target data set is $T_{1}$.
The influenza virus itself, however, may have changed or evolved and the previously built model may not be directly applicable. 
So a RECaST transfer learning model is used to calibrate the source model to this year's target data.
The following year's flu season will see more patients -- this target data set is $T_{2}$.
The online RECaST method uses the source model and the $T_{1}$ model to build a model on the $T_{2}$ patients.
It is possible, however, that the new flu strain is very different between $T_{1}$ and $T_{2}$ and that including this extra information would be harmful.
This would be an example of negative transfer.
Our online method incorporates source and previous target data sets while reducing the influence of negative transfer.
 
The online RECaST method can be applied to the univariate and multivariate outcome cases.
Take $\by_{T_{1}, 1}, \dots \by_{T_{1}, n_{T_{1}}}$ to be a sample of $n_{T_{1}}$ outcomes from the first target, $T_{1}$, and $\by_{T_{2}, 1}, \dots \by_{T_{2}, n_{T_{2}}}$ to be a sample of $n_{T_{2}}$ outcomes from the second target, $T_{2}$, with associated features $\bx_{T_{2}, 1}, \dots \bx_{T_{2}, n_{T_{2}}}$.
Take $\bOmega$ to be the unknown parameters: in the multivariate Cauchy model $\bOmega = \{ \bdelta, \bGamma , \bSigma\}$ and in the multivariate Gaussian copula model $\bOmega = \{\delta_{1}, \dots , \delta_{m} , \gamma_{1} , \dots , \gamma_{m} , \bR , \bSigma \}$.
Let $f(\widehat{ \bTheta }_{S}, \bx)$ be the fitted source model and $p(\bOmega \mid \by_{T_{1}, 1} , \dots \by_{T_{1}, n_{T_{1}}} , \widehat{ \bTheta }_{S})$ be the learned RECaST posterior distribution for $T_{1}$ given by either Equation \eqref{equation: mv cauchy posterior} or Equation \eqref{equation: copula posterior}.
Only the learned posterior distribution on $T_{1}$ is shared and not the data themselves.
In transferring from $S$ and $T_{1}$ to $T_{2}$, we construct a new prior distribution on $\bOmega$ given by
\begin{align*}
\pi_{\star}(\bOmega, \alpha) = \pi(\alpha) \Big\{ \alpha \cdot p(\bOmega \mid \by_{T_{1}, 1} , \dots \by_{T_{1}, n_{T_{1}}} , \widehat{ \bTheta }_{S}) + (1-\alpha) \cdot \pi(\bOmega) \Big\}.
\end{align*}
where $\alpha \in [0,1]$ is a weight parameter to be estimated.
The prior distribution is a convex combination of the RECaST posterior distribution learned from $T_{1}$ and another prior $\pi(\bOmega)$ that is uninformed by $T_{1}$.
The weight parameter $\alpha$ controls the degree to which each prior influences the posterior.
An additional prior $\pi(\alpha)$ is included as a prior on the weight parameter.
A larger $\alpha$ value indicates an increase in the weight of the posterior from $T_{1}$; this is the behavior we would expect the model to learn if $T_{1}$ is informative for $T_{2}$.
If $\alpha = 0$ no information from the $T_{1}$ is used; then this reverts to the same posterior as single target RECaST and avoids negative transfer.
The posterior distribution of $(\bOmega, \alpha)$ is
\begin{align*}
p(\bOmega, \alpha \mid \by_{T_{2}, 1}, \dots \by_{T_{2}, n_{T_{2}}} , \widehat{ \bTheta }_{S}) \propto \pi_{\star}(\bOmega, \alpha) \cdot \prod_{i=1}^{n_{T_{2}}} \int_{\mathbb{R}^{m}} p(\by_{T_{2}, i} \mid \text{diag}(\bbeta_{i}), \bOmega , \widehat{ \bTheta }_{S}) \cdot p(\beta_{i} \mid \bOmega) \ d\bbeta_{i}.
\end{align*}
As this new parameter $\alpha$ in only present in the prior, the posterior predictive sampling algorithms for $\widetilde{ \by }_{T_{2}}$ are unchanged.

The marginal mean of $\alpha$ gives insight into the amount of information shared from $T_{1}$.
First, the marginal posterior distribution of $\alpha$ is
\begin{align*}
p(\alpha & \mid \by_{T_{2}, 1}, \dots \by_{T_{2}, n_{T_{2}}} , \widehat{ \bTheta }_{S}) \\
    & = \int p(\bOmega, \alpha \mid \by_{T_{2}, 1}, \dots \by_{T_{2}, n_{T_{2}}} , \widehat{ \bTheta }_{S}) \ d \bOmega \\
    & \propto \pi(\alpha) \cdot \Bigg\{ 
        \alpha \cdot \Big[ \int p(\bOmega \mid \by_{T_{1}, 1} , \dots \by_{T_{1}, n_{T_{1}}} , \widehat{ \bTheta }_{S}) \cdot p(\by_{T_{2}, 1}, \dots \by_{T_{2}, n_{T_{2}}} \mid \bOmega , \widehat{ \bTheta }_{S}) \ d\bOmega  \Big] +\\
    & \hspace{2cm} (1-\alpha) \cdot \Big[ \int \pi(\bOmega) \cdot p(\by_{T_{2}, 1}, \dots \by_{T_{2}, n_{T_{2}}} \mid \bOmega , \widehat{ \bTheta }_{S}) \ d\bOmega  \Big]
        \Bigg\}.
\end{align*}
The first integral contains information from the $T_{1}$ posterior distribution whereas the second does not.
As they are constant with respect to $\alpha$, take the first integral to be $k_{1}$ and the second to be $k_{2}$.
The marginal mean of $\alpha$ is 
\begin{align*}
\mathbb{E}(\alpha & \mid \by_{T_{2}, 1}, \dots \by_{T_{2}, n_{T_{2}}}) 
    & = \frac{k_{1} \Big\{\mathbb{E}_{\pi(\alpha)}(\alpha)^{2} + \text{Var}_{\pi(\alpha)}(\alpha)\Big\} + k_{2} \Big\{ \mathbb{E}_{\pi(\alpha)}(\alpha) - \mathbb{E}_{\pi(\alpha)}(\alpha)^{2} - \text{Var}_{\pi(\alpha)}(\alpha) \Big\}}
   {k_{1} \mathbb{E}_{\pi(\alpha)}(\alpha) + k_{2} \{ 1 - \mathbb{E}_{\pi(\alpha)}(\alpha) \}},
\end{align*}
where $\mathbb{E}_{\pi(\alpha)}(\alpha)$ and $\text{Var}_{\pi(\alpha)}(\alpha)$ are the expected value and variance of $\alpha$ with respect to its prior distribution, respectively.
From this, we can determine the behavior based on the choice of prior on $\alpha$.
A natural, uninformative choice of prior for $\alpha$ is the $\text{Uniform}(0,1)$, which gives the posterior mean
\begin{align*}
\mathbb{E}(\alpha \mid \by_{T_{2}, 1}, \dots \by_{T_{2}, n_{T_{2}}} ) & = \frac{\frac{2}{3}k_{1} + \frac{1}{3} k_{2}}{k_{1} + k_{2}}.
\end{align*}
While the Metropolis-Hastings algorithm used to estimate the posterior distribution of $(\bOmega, \alpha)$ will explore the entire parameter space, this choice of prior gives a maximum marginal mean of $\alpha$ of $2/3$ if the $T_{1}$ data are informative to learning the posterior on $T_{2}$ or a minimum of $1/3$ if the $T_{1}$ data are not informative.
The value of the posterior mean of $\alpha$ is thus determined by both the similarity between $T_{1}$ and $T_{2}$ and the prior $\pi(\alpha)$.
A prior such as $\text{Beta}(1/a, 1/a)$ for $a > 1$ will give posterior means closer to 0 and 1 for large values of $a$.
This may be preferable in practice if the $T_{1}$ information is very likely (or unlikely) to be informative for $T_{2}$.

This idea naturally extends to more target data sets as well.
If there are $\ell$ target data sets, we take $\balpha = \{\alpha_{1}, \dots , \alpha_{\ell} \}$ to be a weight vector where $\alpha_{k} \geq 0$ for $k = 1 ,\dots , \ell$ and $\sum_{k=1}^{\ell} \alpha_{k} = 1$.
The online posterior distribution for $T_{\ell}$ is given by
\begin{align*}
p & (\bOmega, \balpha \mid \by_{T_{\ell}, 1}, \dots \by_{T_{\ell}, n_{T_{\ell}}} , \widehat{ \bTheta }_{S}) \\
& \propto \pi(\balpha) \Big\{ \alpha_{\ell} \cdot \pi(\bOmega) + \sum_{k = 1}^{\ell - 1} \alpha_{k} \cdot p(\bOmega \mid \by_{T_{k}, 1} , \dots \by_{T_{k}, n_{T_{k}}} , \widehat{ \bTheta }_{S})\Big\} \cdot \\
& \hspace{0.5cm}  \prod_{i=1}^{n_{T_{\ell}}} \int_{\mathbb{R}^{m}} p(\by_{T_{\ell}, i} \mid \bbeta_{i} , \widehat{ \bTheta }_{S}) \cdot p(\bbeta_{i} \mid \bOmega) \ d\bbeta_{i},
\end{align*}
where $\pi(\balpha)$ is now a multivariate prior on $\balpha$, for example a Dirichlet distribution.
An analogous derivation for the posterior means of the elements of $\balpha$ for the multivariate prior is provided in Appendix \ref{section: online mean derivation}.

\section{Simulation Study}
\label{section: simulation study}

\subsection{Overview}
\label{section: simulation study overview}

In this section we examine the performance of the multivariate and online RECaST methods through simulations on synthetic data.
We consider two methods for comparison.
The first is a group ridge regression model that is built only on the target data set implemented with the \texttt{glmnet} package in \texttt{R} \citep{JSSv033i01}.
The second is the univariate RECaST method where a posterior is fit on each outcome separately.
This will serve as validation to determine whether the multivariate RECaST methods improve predictive performance and uncertainty quantification.

The source model for the RECaST Cauchy and RECaST copula methods is a group ridge regression fit to the source data.
The Mahalanobis distance $d(\widetilde{ \by }, \widehat{ \by }) = \sqrt{ (\widetilde{ \by } - \widehat{ \by })^{\top} \bS^{-1} (\widetilde{ \by } - \widehat{ \by })}$ measures the distance between the test values $\widetilde{ \by }$ and the predicted values $\widehat{ \by }$, where $\bS$ is the empirical covariance of the test set.
The test sets are of size 100.
The posterior predictive sampling values are $n_{\text{post}} = 50$, $n_{\bB} = 50$, and $n_{\bY} = 20$, and the posterior predictive sample is of size $50,000$ for each test observation.
This sample is used to calculate the empirical coverage and the Mahalanobis distance.
Each simulation setting is repeated over 100 target test data sets.
The Mahalanobis distances and empirical coverage values reported are averages the over the 100 target test data sets.

The source sample size is large at $n_{S} = 1,000$. The data are generated from a multivariate linear regression with fixed source data generating parameters $\bTheta_{S}$ with $p = 50$ features (including an intercept) and $m = 2$ responses, $\bY_{S} \sim \mathcal{N}_{p}(\bTheta_{S} \bx_{S}, \bSigma)$ where $\bSigma$ is the covariance between the responses.
The features are generated from the standard Gaussian distribution, $\bx_{S} \sim \mathcal{N}_{p-1}(\bzero, \bI_{p-1})$.

\subsection{Multivariate Single Target Simulations}
\label{section: multivariate simulations}

While the source sample size $n_{S} = 1,000$ is fixed, the target sample sizes vary: $n_{T} \in \{ 20, 50, 100\}$.
With the data generating mechanism of linear regression, the ridge baseline model built only on the target data should perform well with 100 target data points.
This provides a good test for negative transfer since the ridge baseline should have enough data to reliably estimate the true data generating parameters.
With a sample size of 50, there are as many data points in the target as there are features and for a sample size of 20, there are fewer data points than features.
With such little data, it is expected that target only models would perform poorly and that the inclusion of source information could be beneficial.
We also vary the covariance between the outcomes $\bSigma$ used to generate data: no covariance $\bSigma = \big[ \begin{smallmatrix}
1 & 0 \\ 0 & 1
\end{smallmatrix} \big]$, positive covariance $\bSigma = \big[  \begin{smallmatrix}
1 & 0.5 \\ 0.5 & 1
\end{smallmatrix}\big]$, and negative covariance $\bSigma = \big[ \begin{smallmatrix}
1 & -0.5 \\ -0.5 & 1
\end{smallmatrix} \big]$.

We first consider an additive relationship between the source and target data generating parameters.
The source parameters for the first response $\bTheta_{S, 1,\cdot}$ are fixed between 0.5 and 5 and the parameters for the second response $\bTheta_{S, 2, \cdot}$ are fixed between $-0.5$ and $-5$.
The target data generating parameters are constructed by adding uniformly distributed noise to $\bTheta_{S}$.
The target parameters for the first response are generated as $\bTheta_{T, 1, \cdot} = \bTheta_{S, 1, \cdot} + U(0,a)$ and the parameters for the second response are generated as $\bTheta_{T, 2, \cdot} = \bTheta_{S,2,\cdot} + U(b, 0)$.
This controls the similarity between the source and the target data: as $a$ increases or $b$ decreases, the target data generating parameters can grow further from the source parameters.
We consider values of $(a,b) \in \{ (0.5, -0.5), (1, -1), (2, -2)\}$.

Tables \ref{table: mv recast shift sigma 1}, \ref{table: mv recast shift sigma 2}, and \ref{table: mv recast shift sigma 3} summarize the performance results stratified by $\bSigma$.
RECaST using a multivariate Cauchy and a multivariate Gaussian copula is presented as MV Cauchy and MV Copula, respectively.
In all cases, the multivariate RECaST methods have better predictive performance than the univariate method, demonstrating the benefit of considering the outcomes jointly.
The multivariate methods also boast lower standard errors than univariate RECaST.
This shows that the multivariate methods have more concentrated predictions.
Across every scenario, both multivariate RECaST methods outperform the baseline ridge when $n_{T} = 20 \text{ or } 50$ as shown by the lower Mahalanobis distances.
As expected, as the target sample size increases the baseline ridge method improves.
In fact, ridge has lower Mahalanobis distances when $n_{T} = 100$, indicating negative transfer.
Even with a large sample size, when the source and target data sets are similar, $(a,b) = (0.5, -0.5)$, the Mahalanobis distances for the ridge and RECaST methods are close (in some cases less than a standard error apart) indicating less negative transfer.
In contrast, when the sample size is large ($n_{T} = 100$) and the source and target are the most dissimilar, $(a,b) = (2,-2)$, negative transfer is most apparent.

There is slight improvement in performance when the covariance between the outcomes is negative compared to when they are independent shown by the decrease in Mahalanobis distances.
There is a similar slight decrease in performance when the responses are positively correlated.
For every scenario, the multivariate RECaST methods achieve excellent empirical coverage of 96\% or 97\% at the 95\% nominal level.
The marginal coverages from univariate RECaST are often 100, which is a result of the entire outcome space being in the credible set.
In these cases the credible set is uninformative at the 95\% level.

\begin{table}[H]
    \caption{Performance metrics for $\bTheta_{T,1,\cdot} = \bTheta_{S,1,\cdot} + U(0,a)$ and $\bTheta_{T,2,\cdot} = \bTheta_{S,2,\cdot} + U(b,0)$ with $\bSigma =\Big[ \begin{smallmatrix}
1 & 0 \\
0 & 1 
\end{smallmatrix} \Big]
 $. The reported values are averaged over 100 source and target data sets: average Mahalanobis distance (standard error of Mahalanobis distances) [empirical coverage at the 95\% nominal level].
    For univariate RECaST the marginal coverages are reported.}
    \centering
    \begin{tabular}{ c c | c c c c}
     $a, \ b$ & $n_T$ & Ridge & Univariate & MV Cauchy & MV Copula\\ \hline

   $0.5, -0.5$ & 20 & 1.2 (0.23) & 1.1 (0.19) [100, 100] & 0.18 (0.032) [97] & 0.18 (0.032) [97] \\
      & 50 & 0.48 (0.09)  & 1.1 (0.14) [100, 100] & 0.17 (0.022) [97] & 0.17 (0.022) [97] \\
     & 100 & 0.21 (0.02) & 1 (0.1) [99, 99] & 0.16 (0.016) [97]& 0.16 (0.015) [97] \\ \hline
         $1, -1$& 20 & 1.2 (0.24) & 1.1 (0.2) [99, 100] & 0.29 (0.051) [97] & 0.29 (0.052) [97] \\
        & 50 & 0.47 (0.09) & 1.1 (0.15) [100, 100] & 0.27 (0.039) [97] & 0.27 (0.039) [97]\\
     & 100 & 0.2 (0.02) & 1 (0.11) [100, 100] & 0.27 (0.032) [97]& 0.27 (0.031) [97] \\ \hline
         $2, -2$ & 20 & 1.2 (0.25) & 1.1 (0.21) [99, 100]& 0.5 (0.096) [97] & 0.49 (0.09) [97]  \\
          & 50 & 0.44 (0.09) & 1.1 (0.15) [100, 100] & 0.48 (0.078) [97] & 0.47 (0.078) [97]  \\
         &  100 & 0.18 (0.02) & 1.1 (0.11) [100, 100]  & 0.47 (0.067) [97] & 0.46 (0.065) [97]
    \end{tabular}

    \label{table: mv recast shift sigma 1}

\end{table}

\begin{table}[H]
    \caption{Performance metrics for $\bTheta_{T,1,\cdot} = \bTheta_{S,1,\cdot} + U(0,a)$ and $\bTheta_{T,2,\cdot} = \bTheta_{S,2,\cdot} + U(b,0)$ with $\bSigma =\Big[ \begin{smallmatrix}
1 & 0.5 \\
0.5 & 1 
\end{smallmatrix} \Big]
 $. The reported values are averaged over 100 source and target data sets: average Mahalanobis distance (standard error of Mahalanobis distances) [empirical coverage at the 95\% nominal level].
    For univariate RECaST the marginal coverages are reported.}
    \centering
    \begin{tabular}{ c c | c c c c}
     $a, \ b$ & $n_T$ & Ridge & Univariate & MV Cauchy & MV Copula\\ \hline

	 $0.5, -0.5$ & 20 & 1.2 (0.23) & 1.1 (0.19) [100, 100] & 0.19 (0.035) [97]& 0.19 (0.035) [97]\\
	& 50 & 0.42 (0.09) & 1.1 (0.14) [99, 100]  & 0.18 (0.023) [97] & 0.18 (0.023) [97]\\
	& 100 & 0.16 (0.02) & 1 (0.1) [99, 99] & 0.18 (0.017) [97]& 0.18 (0.017) [97]\\ \hline
	$1, -1$& 20 & 1.2 (0.24) & 1.1 (0.2) [99, 100] & 0.3 (0.052) [97] & 0.29 (0.051) [97]\\
	& 50 & 0.42 (0.09) & 1.1 (0.14) [100, 100] & 0.28 (0.04) [97]& 0.28 (0.039) [97]\\
	& 100 & 0.15 (0.02)  & 1 (0.1) [100, 100] & 0.27 (0.032) [96]& 0.27 (0.031) [96] \\ \hline
	$2, -2$ & 20 & 1.2 (0.25) & 1.1 (0.19) [100, 100]  & 0.5 (0.095) [96] & 0.5 (0.095) [97]\\
	& 50  & 0.4 (0.09) & 1.1 (0.14) [100, 100] & 0.48 (0.078) [97]& 0.48 (0.078) [97]\\ 
	& 100 & 0.14 (0.02) & 1.1 (0.11) [100, 100] & 0.47 (0.067) [97]& 0.47 (0.067) [97]
\end{tabular}
    \label{table: mv recast shift sigma 2}

\end{table}

\begin{table}[H]
    \caption{Performance metrics for $\bTheta_{T,1,\cdot} = \bTheta_{S,1,\cdot} + U(0,a)$ and $\bTheta_{T,2,\cdot} = \bTheta_{S,2,\cdot} + U(b,0)$ with $\bSigma =\Big[ \begin{smallmatrix}
1 & -0.5 \\
-0.5 & 1 
\end{smallmatrix} \Big]
 $. The reported values are averaged over 100 source and target data sets: average Mahalanobis distance (standard error of Mahalanobis distances) [empirical coverage at the 95\% nominal level].
    For univariate RECaST the marginal coverages are reported.}
    \centering
    \begin{tabular}{ c c | c c c c}
     $a, \ b$ & $n_T$ & Ridge & Univariate & MV Cauchy & MV Copula\\ \hline

 $0.5, -0.5$ & 20  & 1.2 (0.23) & 1.1 (0.19) [100, 100] & 0.17 (0.03) [97]& 0.16 (0.03) [97]\\
     & 50 & 0.54 (0.09) & 1.1 (0.14) [99, 100]& 0.16 (0.02) [97]& 0.15 (0.019) [97] \\
      & 100 & 0.24 (0.03) & 1 (0.1) [99, 99]& 0.15 (0.014) [97]& 0.15 (0.014) [97] \\ \hline
     $1, -1$& 20 & 1.2 (0.24) & 1.1 (0.2) [99, 99]  & 0.28 (0.051) [97] & 0.28 (0.052) [97]\\
     & 50 & 0.525 (0.09) & 1.1 (0.15) [100, 100] & 0.27 (0.038) [97] & 0.26 (0.037) [97]\\
     & 100 & 0.24 (0.03) & 1 (0.11) [100, 100] & 0.26 (0.032) [97]& 0.26 (0.031) [97]\\ \hline
     $2, -2$ & 20 & 1.2 (0.25) & 1.1 (0.2) [99, 99]& 0.46 (0.067) [97] & 0.46 (0.065) [97] \\
      & 50 & 0.48 (0.09)  & 1.1 (0.15) [100, 100] & 0.47 (0.078) [97]  & 0.47 (0.077) [97] \\
     & 100  & 0.21 (0.03) & 1.1 (0.11) [100, 100]  & 0.46 (0.067) [97]  & 0.46 (0.065) [97]
    \end{tabular}

    \label{table: mv recast shift sigma 3}

\end{table}

We next consider a multiplicative relationship between the source and target data generating parameters.
Our source parameters remain the same as above, but now the target parameters are constructed as $\bTheta_{T} = c \cdot \bTheta_{S}$ with $c \in \{ 0.5, 2 \}$.
The data are generated as above.
Table \ref{table: mv recast scale} summarizes the results.
As before, the baseline ridge regression performance improves as the sample size increases; however, notice that for $n_{T} = 100$ both multivariate RECaST methods have lower Mahalanobis distances for both $c = 0.5$ and $c = 2$ than the baseline ridge.
In this setting, the multivariate RECaST methods are robust to negative transfer.
They also outperform univariate RECaST.
Here we see a greater improvement in predictive performance for both multivariate RECaST methods when the results are positively correlated than in the previous simulation.
Both methods again provide excellent coverage slightly above the 95\% nominal level.

\begin{table}[H]
    \caption{Performance metrics for $\bTheta_{T} = c \cdot \bTheta_{S}$. The reported values are averaged over 100 source and target data sets: average Mahalanobis distance (standard error of Mahalanobis distances) [empirical coverage at the 95\% nominal level].
    For univariate RECaST, the marginal coverages are reported.}
    \centering
    \begin{tabular}{ c c c | c c c c }
       $\bSigma$ & $c$ & $n_T$ & Ridge & Univariate & MV Cauchy & MV Copula \\ \hline
$
\Big[ \begin{smallmatrix}
	1 & 0 \\
	0 & 1 
\end{smallmatrix}\Big]
$
& 0.5 & 20 & 1.2 (0.23) & 1.1 (0.21) [100, 100] & 0.23 (0.046) [97]& 0.22 (0.045) [97]\\
& & 50 & 0.74 (0.12) & 1 (0.13) [99, 99]  & 0.21 (0.022) [97] & 0.21 (0.022) [97]\\
&  & 100 & 0.39 (0.04)  & 1 (0.1) [98, 99] & 0.2 (0.02) [97]& 0.2 (0.019) [97] \\ \hline

&  2 & 20 & 1.2 (0.24) & 1 (0.2) [100, 100] & 0.06 (0.012) [97]& 0.06 (0.011) [97] \\
&  & 50 & 0.36 (0.08) & 1 (0.13) [99, 99] & 0.06 (0.0061) [97]& 0.06 (0.0061) [97]\\
& &  100  & 0.11 (0.012) & 1 (0.1) [98, 99] & 0.06 (0.0053) [96]& 0.06 (0.0053) [97]\\ \hline \hline 

$
\Big[ \begin{smallmatrix}
	1 & 0.5 \\
	0.5 & 1 
\end{smallmatrix} \Big]
$
& 0.5& 20 & 1.2 (0.24) & 1 (0.21) [100, 100] & 0.25 (0.051) [97] & 0.25 (0.053) [97] \\
& & 50  & 0.6 (0.1) & 1 (0.13) [99, 99]  & 0.23 (0.026) [97] & 0.23 (0.026) [97] \\
&  & 100 & 0.3 (0.03) & 1 (0.1) [98, 99]  & 0.23 (0.024) [97] & 0.23 (0.024) [97] \\ \hline
& 2& 20  & 1.2 (0.24)  & 1 (0.2) [100, 100] & 0.07 (0.013) [97] & 0.07 (0.0073) [97]\\
& & 50 & 0.34 (0.08)  & 1 (0.13) [99, 99] & 0.07 (0.0073) [97]& 0.07 (0.0073) [97] \\
& & 100 & 0.09 (0.01) & 1 (0.1) [98, 99] & 0.07 (0.0064) [97] & 0.07 (0.0065) [97]
\end{tabular}

    \label{table: mv recast scale}

\end{table}

\subsection{Multivariate Online Simulations}
\label{section: multivariate online simulations}

We now consider two target data sets, $T_{1}$ and $T_{2}$. 
The data are again generated from multivariate linear regressions with $p = 50$ features (including an intercept) and $m = 2$ outcomes.
The source data generating parameters are fixed within the intervals $\bTheta_{S,1,\cdot} \in [2, 2.5]$ and $\bTheta_{S, 2, \cdot } \in [-2.5, -2]$.
The source sample size is $n_{S} = 1,000$.
The target $T_{1}$ data generating parameters $\bTheta_{T_{1}}$ are also fixed within the intervals $\bTheta_{T_{1}, 1, \cdot} \in [6.5, 7]$ and $\bTheta_{T_{1}, 2, \cdot} \in [-7, -6.5]$.
We take the target $T_{1}$ sample size to be large with $n_{T_{1}} = 1,000$.
Thus, this analysis isolates the effects of the online results on target $T_{2}$.

As before, we use sample sizes of $n_{T_{2}} \in \{ 20, 50, 100 \}$.
We test the same covariance structures as in Section \ref{section: multivariate simulations} with the responses either having no covariance, positive covariance, or negative covariance.
The data generating parameters for the domain of interest, $T_{2}$, are based on the other target data set $T_{1}$.
We take $\bTheta_{T_{2},1,\cdot} = \bTheta_{T_{1}, 1, \cdot} + U(0, a)$ and $\bTheta_{T_{2},2,\cdot} = \bTheta_{T_{1}, 2, \cdot} + U(b, 0)$.
We consider two scenarios, the first with $a = 0$ and $b = 0$ and the second with $a = 3$ and $b = -3$.
The first corresponds to the $T_{1}$ and $T_{2}$ data coming from close underlying distributions.
In this case, we expect the multivariate, online methods to perform well because the posterior distribution from $T_{1}$ should be informative.
In the second scenario, the target data sets are not coming from close distributions as the data generating parameters could be substantially different.

We compare the multivariate RECaST methods to a baseline group ridge model fit only on the $T_{2}$ data and to univariate RECaST.
For all RECaST methods, we use a group ridge model fit on the source data as the source model.
The multivariate RECaST methods we consider in this analysis are: 

\begin{itemize}
    \item multivariate RECaST using the multivariate Cauchy distribution learned on $T_{2}$ (MV Cauchy),
    \item multivariate, online RECaST using a multivariate Cauchy distribution learned on $T_{2}$ with a prior informed by the posterior fit on $T_{1}$ (MV-On Cauchy),
    \item multivariate RECaST using the multivariate Normal copula learned on $T_{2}$ (MV Copula),
    \item multivariate, online RECaST using a multivariate Normal copula learned on $T_{2}$ with a prior informed by the posterior fit on $T_{1}$ (MV-On Copula).
\end{itemize}
A standard uniform distribution is used as the prior on the weight parameter $\alpha$.

Tables \ref{table: mvmt sim mahalanobis sigma 1}, \ref{table: mvmt sim mahalanobis sigma 2}, and \ref{table: mvmt sim mahalanobis sigma 3} summarize the predictive performance results for the multivariate, online simulations stratified by $\bSigma$.
All of the multivariate RECaST methods have better predictive performance univariate RECaST.
They also had lower Mahalanobis distances than the ridge baseline when $n_{T_{2}} = 20$.
This is also the case when $n_{T_{2}} = 50$ except when there is negative covariance between the outcomes and $\bTheta_{T_{2}}$ if far from $\bTheta_{T_{1}}$.
When $n_{T_{2}} = 100$ the effects of negative transfer are clear as the baseline ridge method outperforms all RECaST methods.

We do not see any effects of negative transfer between $T_{1}$ and $T_{2}$, even when $\bTheta_{T_{2}}$ if far from $\bTheta_{T_{1}}$.
This is evident by comparing the MV Cauchy and MV-On Cauchy Mahalanobis distances and the MV Copula and MV-On Copula Mahalanobis distances.
In both cases the MV-On method performs as well or better, indicated by an equal or smaller Mahalanobis distance, than the corresponding offline method. 
This is even the case when $\bTheta_{T_{2}}$ if far from $\bTheta_{T_{1}}$.
This lends merit to the idea of using a convex combination of priors to mitigate negative transfer.
The MV-On Cauchy has the best performance shown by the lowest Mahalanobis distances.
This difference between methods is most apparent when the $T_{2}$ sample sizes are small; this is where transfer learning would be the most necessary as there is the least amount of information provided in the target domain of interest.

Tables \ref{table: mvmt sim coverage sigma 1}, \ref{table: mvmt sim coverage sigma 2}, and \ref{table: mvmt sim coverage sigma 3} present the empirical coverage values at the 95\% nominal level for the RECaST methods, stratified by $\bSigma$.
In all cases, the MV Cauchy, MV-On Cauchy, and MV Copula RECaST methods achieve empirical coverage of at least 95\%.
This is true even when the $\bTheta_{T_{2}}$ is far from $\bTheta_{T_{1}}$.
The MV-On Copula method under-covers slightly when sample sizes ar larger, but achieves nominal coverage when transfer learning is most necessary, when $n_{T_{2}} = 20$.
In most cases, univariate RECaST provides marginal coverages of 100, indicating that the credible sets are not helpful at the 95\% nominal level.

For the online methods, Tables \ref{table: mvmt alpha sigma 1}, \ref{table: mvmt alpha sigma 2}, and \ref{table: mvmt alpha sigma 3} summarize the average estimated $\alpha$ weights, stratified by $\bSigma$.
For the multivariate Cauchy method, the posterior mean of $\alpha$ is centered around $0.67$ when $\bTheta_{T_{2}}$ is close to $\bTheta_{T_{1}}$.
Since a uniform distribution was used as the prior for $\alpha$, this is the expected mean when the posterior for $T_{1}$ is more useful than an uninformative prior.
When $\bTheta_{T_{2}}$ is far from $\bTheta_{T_{1}}$, more information is shared when sample sizes are small.
As $n_{T_{2}}$ increases, the information from $T_{1}$ does not need to be relied upon as much and the posterior mean of $\alpha$ decreases.
The copula-based method tends to borrow the same amount of information from $T_{1}$ in all settings except when there is negative correlation between the outcomes.

\begin{table}[H]
\scriptsize
\centering 
\caption{Mahalanobis distances for multivariate, online RECaST averaged over 100 source and target data sets for $\bSigma = \Big[ \begin{smallmatrix}
1  & 0 \\
0  & 1 
\end{smallmatrix} \Big]$. The reported values are: Average Mahalanobis distance (standard error of Mahalanobis distances)}
\begin{tabular}{ c c | c c c c c c}
 $a, b$ 
 & $n_{T_{2}}$ 
 & Ridge 
 & Univariate 
 & MV Cauchy 
 & MV-On Cauchy 
 & MV Copula 
 & MV-On Copula\\ \hline
 0, 0 
 & 20 
 &  1.2 (0.23) 
 & 1 (0.20) 
 & 0.67 (0.16) 
 & 0.62 (0.11) 
 & 0.65 (0.13) 
 & 0.64 (0.12) 
 \\
 
 & 50 
 & 0.94 (0.15) 
 & 0.96 (0.14) 
 & 0.61 (0.08) 
 & 0.59 (0.08) 
 & 0.59 (0.08) 
 & 0.59 (0.08) 
 \\

 & 100 
 & 0.52 (0.06)  
 & 0.92 (0.10) 
 & 0.58 (0.07) 
 & 0.58 (0.07) 
 & 0.61 (0.18) 
 & 0.61 (0.19) 
 \\ \hline
 $3, -3$ 
 & 20 & 1.2 (0.26)  & 12 (2.6)& 0.97 (0.22) & 0.93 (0.18) & 1.04 (0.26) & 1.03 (0.25)

 \\
 
 & 50 & 0.46 (0.1)& 11 (1.7)& 0.87 (0.11)& 0.87 (0.11) & 0.86 (0.11) & 0.86 (0.11)

 \\
 
 & 100 & 0.18 (0.03) & 11 (1.3)  & 0.83 (0.1) & 0.83 (0.1) & 0.83 (0.1) & 0.83 (0.1)

\end{tabular}
\label{table: mvmt sim mahalanobis sigma 1}
\end{table}

\begin{table}[H]
\scriptsize
\centering 
\caption{Mahalanobis distances for multivariate, online RECaST averaged over 100 source and target data sets for $\bSigma = \Big[ \begin{smallmatrix}
1  & 0.5 \\
0.5  & 1 
\end{smallmatrix} \Big]$. The reported values are: Average Mahalanobis distance (standard error of Mahalanobis distances)}
\begin{tabular}{ c c | c c c c c c}
 $a, b$ 
 & $n_{T_{2}}$ 
 & Ridge 
 & Univariate 
 & MV Cauchy 
 & MV-On Cauchy 
 & MV Copula 
 & MV-On Copula\\ \hline
 0, 0 
 & 20 
 & 1.3 (0.22)  
 & 1 (0.19) 
 & 0.71 (0.13) 
 & 0.67 (0.11) 
 & 0.7 (0.12) 
 & 0.7 (0.13) 
 \\
 
 & 50  
 & 1.1 (0.16)  
 & 0.95 (0.14) 
 & 0.67 (0.1) 
 & 0.66 (0.09) 
 & 0.66 (0.09) 
 & 0.66 (0.09) 
 \\
 
 & 100 
 & 0.62 (0.07)  
 & 0.93 (0.11) 
 & 0.64 (0.07) 
 & 0.63 (0.07) 
 & 0.64 (0.07) 
 & 0.64 (0.08) 
 \\ \hline
 $3, -3$ 
 & 20  & 1.2 (0.26)  & 12 (2.5) & 0.98 (0.23) & 0.93 (0.19) & 1.02 (0.24) & 1.02 (0.23)

 \\
 
 & 50  & 0.51 (0.1) & 11 (1.6) & 0.86 (0.11) & 0.86 (0.12) & 0.86 (0.11) & 0.86 (0.11)

 \\
 
 & 100  & 0.22 (0.03) & 11 (1.3) & 0.84 (0.1) & 0.84 (0.1) & 0.83 (0.1) & 0.83 (0.1)

\end{tabular}
\label{table: mvmt sim mahalanobis sigma 2}
\end{table}

\begin{table}[H]
\scriptsize
\centering 
\caption{Mahalanobis distances for multivariate, online RECaST averaged over 100 source and target data sets for $\bSigma = \Big[ \begin{smallmatrix}
1  & -0.5 \\
-0.5  & 1 
\end{smallmatrix} \Big]$. The reported values are: Average Mahalanobis distance (standard error of Mahalanobis distances)}
\begin{tabular}{ c c | c c c c c c}
 $a, b$ 
 & $n_{T_{2}}$ 
 & Ridge 
 & Univariate 
 & MV Cauchy 
 & MV-On Cauchy 
 & MV Copula 
 & MV-On Copula\\ \hline
0, 0 
 & 20  
 & 1.2 (0.24)  
 & 1 (0.20) 
 & 0.61 (0.13) 
 & 0.56 (0.1) 
 & 0.58 (0.12) 
 & 0.58 (0.12) 
 \\
 
 & 50  
 & 0.76 (0.13)  
 & 0.95 (0.14) 
 & 0.53 (0.07) 
 & 0.53 (0.07) 
 & 0.53 (0.07) 
 & 0.52 (0.07) 
 \\
 
 & 100 
 & 0.4 (0.04)  
 & 0.92 (0.1) 
 & 0.53 (0.05) 
 & 0.52 (0.05) 
 & 0.53 (0.09) 
 & 0.51 (0.11) 
\\ \hline
 $3, -3$ 
 & 20  & 1.2 (0.26) & 12 (2.6) & 0.98 (0.22) & 0.92 (0.18) & 1.04 (0.27) & 1.04 (0.27)

 \\
 
 & 50  & 0.4 (0.1) & 11 (1.7) & 0.87 (0.12) & 0.87 (0.12) & 0.86 (0.11) & 0.86 (0.11)

 \\
 
 & 100  & 0.14 (0.02) & 11 (1.3) & 0.83 (0.1) & 0.83 (0.1) & 0.83 (0.1) & 0.83 (0.1)

 \\  
\end{tabular}
\label{table: mvmt sim mahalanobis sigma 3}
\end{table}

\begin{table}[H]
\centering 
\small
\caption{Empirical coverage values at the 95\% nominal level for multivariate, online RECaST and empirical coverage values averaged over 100 source and target data sets for $\bSigma = \Big[ \begin{smallmatrix}
    1 & 0 \\ 0 & 1
\end{smallmatrix} \Big]$. For univariate RECaST the marginal coverages are reported.}
\begin{tabular}{ c c | c c c c c}
$a, b$ 
  & $n_{T_{2}}$ 
  & Univariate 
& MV Cauchy 
& MV-On Cauchy 
& MV Copula 
& MV-On Copula\\ \hline
  
  0, 0 
& 20 
& 100, 100 
& 97
& 97 
& 97 
& 96
\\

  & 50 
& 100, 100 
& 96
& 97 
& 97 
& 94
\\

  & 100 
& 100, 100 
& 96 
& 97
& 97  
& 93
\\ \hline

 $3, -3$ 
& 20 & 96, 95 & 96 & 96 & 97 & 95

\\

  & 50 & 99, 100 & 96 & 97 & 97 & 94

\\

 & 100 & 100, 100 & 97 & 97 & 97 & 93

\end{tabular}
\label{table: mvmt sim coverage sigma 1}
\end{table}

\begin{table}[H]
\centering 
\small
\caption{Empirical coverage values at the 95\% nominal level for multivariate, online RECaST and empirical coverage values averaged over 100 source and target data sets for $\bSigma = \Big[ \begin{smallmatrix}
    1 & 0.5 \\ 0.5 & 1
\end{smallmatrix} \Big]$. For univariate RECaST the marginal coverages are reported.}
\begin{tabular}{ c c | c c c c c}
$a, b$ 
  & $n_{T_{2}}$ 
  & Univariate 
& MV Cauchy 
& MV-On Cauchy 
& MV Copula 
& MV-On Copula\\ \hline
 0, 0 
& 20 
& 100, 100 
& 97 
& 96 
& 96 
& 97
\\

  & 50  
& 100, 100 
& 96
& 97 
& 97 
& 94
\\

  & 100 
& 100, 100 
& 96
& 96 
& 97 
& 94
\\ \hline

 $3, -3$ 
& 20 & 96, 95 & 95 & 96 & 97 & 95

\\

  & 50   & 99, 100 & 96 & 97 & 97 & 94

\\

  & 100 & 100, 100 & 97 & 97 & 97 & 93

\end{tabular}
\label{table: mvmt sim coverage sigma 2}
\end{table}

\begin{table}[H]
\centering 
\small
\caption{Empirical coverage values at the 95\% nominal level for multivariate, online RECaST and empirical coverage values averaged over 100 source and target data sets for $\bSigma = \Big[ \begin{smallmatrix}
    1 & -0.5 \\ -0.5 & 1
\end{smallmatrix} \Big]$. For univariate RECaST the marginal coverages are reported.}
\begin{tabular}{ c c | c c c c c}
$a, b$ 
  & $n_{T_{2}}$ 
  & Univariate 
& MV Cauchy 
& MV-On Cauchy 
& MV Copula 
& MV-On Copula\\ \hline
 0, 0 
& 20  
& 100, 100 
& 96  
& 96 
& 97 
& 96
\\
 
  & 50  
& 100, 100 
& 96
& 97 
& 97 
& 95
\\
 
  & 100 
& 100, 100 
& 97
& 97  
& 97 
& 94
\\ \hline
 $3, -3$ 
& 20  & 95, 95 & 96 & 96 & 97 & 95

\\
 
  & 50 & 99, 100 & 96 & 96 & 97 & 94

\\
 
  & 100 & 100, 100 & 97 & 97 & 97 & 93

\\  
\end{tabular}
\label{table: mvmt sim coverage sigma 3}
\end{table}

\begin{table}[H]
\centering 
\caption{Posterior means of $\alpha$ for multivariate, online RECaST averaged over 100 source and target data sets for $\bSigma = \Big[ \begin{smallmatrix}
1 & 0 \\
0 & 1 
\end{smallmatrix}\Big]$: average posterior mean (standard deviation of the posterior means).}
\begin{tabular}{c c | c c }
$a, b$ & $n_{T_{2}}$ &  MV-On Cauchy & MV-On Copula\\ \hline

   0, 0 & 20 & 0.67 (0.0028) & 0.46 (0.15) \\
 & 50  & 0.67 (0.0027)  & 0.47 (0.15)  \\
 & 100 & 0.67 (0.0031) & 0.45 (0.15) \\ \hline
 $3, -3$ & 20 & 0.62 (0.12) & 0.47 (0.15) \\
 & 50  & 0.41 (0.14)  & 0.47 (0.15)\\
 & 100 & 0.34 (0.057) & 0.46 (0.15)
\end{tabular}
\label{table: mvmt alpha sigma 1}
\end{table}

\begin{table}[H]
\centering 
\caption{Posterior means of $\alpha$ for multivariate, online RECaST averaged over 100 source and target data sets for $\bSigma = \Big[ \begin{smallmatrix}
1 & 0.5 \\
0.5 & 1 
\end{smallmatrix}\Big]$: average posterior mean (standard deviation of the posterior means).}
\begin{tabular}{c c | c c }
$a, b$ & $n_{T_{2}}$ &  MV-On Cauchy & MV-On Copula\\ \hline

 0, 0 & 20 & 0.67 (0.0027) & 0.51 (0.15) \\
 & 50 & 0.67 (0.0026) & 0.47 (0.15) \\
 & 100 & 0.67 (0.0032) & 0.47 (0.15) \\ \hline
 $3, -3$ & 20  & 0.59 (0.14) & 0.47 (0.15) \\
 & 50  & 0.38 (0.12)  & 0.47 (0.14)\\
 & 100 & 0.35 (0.066) & 0.45 (0.14)
\end{tabular}
\label{table: mvmt alpha sigma 2}
\end{table}

\begin{table}[H]
\centering 
\caption{Posterior means of $\alpha$ for multivariate, online RECaST averaged over 100 source and target data sets for $\bSigma = \Big[ \begin{smallmatrix}
1 & -0.5 \\
-0.5 & 1 
\end{smallmatrix}\Big]$: average posterior mean (standard deviation of the posterior means).}
\begin{tabular}{c c | c c }
$a, b$ & $n_{T_{2}}$ &  MV-On Cauchy & MV-On Copula\\ \hline
 0, 0 & 20 & 0.67 (0.0027) & 0.34 (0.11) \\
 & 50 & 0.67 (0.0029) & 0.37 (0.1)  \\
 & 100 & 0.67 (0.0029)  & 0.39 (0.12)
 \\ \hline
 $3, -3$ & 20  & 0.64 (0.091) & 0.38 (0.11) \\
 & 50   & 0.43 (0.15)  & 0.41 (0.13) \\
 & 100  & 0.36 (0.096) & 0.39 (0.12) \\  
\end{tabular}
\label{table: mvmt alpha sigma 3}
\end{table}


\section{Dental Data Analysis}
\label{section: dental data}

Difficult measurements need to be taken at each tooth to diagnose periodontitis.
A reliable model for these outcomes would help alleviate the burden on dentists and provide additional information to inform their decision making.

\begin{figure}[h]
    \centering
    \caption{The outcome measurements for each participant's first visit.}
    \includegraphics[width=0.5\linewidth]{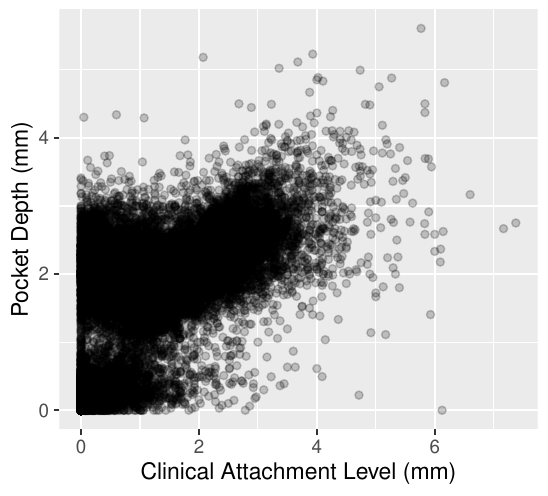}
    \label{fig: HP outcome plot}
\end{figure}

Participants included in the data set had at least 8 years of continuous dental insurance during the study period.
We consider two periodontal measurements measured in millimeters (mm): clinical attachment level (CAL) and pocked depth (PD).
PD is the ``distance from the gingival margin to the base of the gingival sulcus or periodontal pocket" and CAL is the ``distance from the cemento-enamel junction (or another definite chosen landmark) to the base of the sulcus or periodontal pocket" \citep{page2007case}.
These are both important measurements in diagnosing periodontitis.
A general categorization of periodontitis is: {\em slight} = 1 to 2 mm of CAL, {\em moderate} = 3 to 4 mm of CAL, and {\em severe} $\geq$ 5 mm of CAL \citep{periodontitis2015american}.
Figure \ref{fig: HP outcome plot} shows the skewness in the outcome measurements; CAL has a range of [0, 7.75] and PD has a range of [0, 6.2].
These measurements are {\em whole-mouth averages}, averaged over all of each participant's teeth.

Our goal is to be able to predict CAL and PD based on a number of demographic, general health, and insurance variables.
These outcomes are time consuming and challenging to measure; being able to predict them would aid in predicting gum disease.
To achieve this we take the information of each participant at their first visit.
There are 23,529 participants considered.
There is a large disparity in the racial groups represented in the study.
Of those participants 21,029 are White, 1,297 are Black or African American, 1,050 are Asian, and 153 are Native American or Alaskan Native.
Table \ref{table: HP Data} shows the breakdown of the measured features and outcomes by demographic group. 
As we saw in the literature in Section \ref{section: related work}, using models built on a majority population to make predictions on underrepresented groups can produce poor results.
To address this, we use these groups to split the data set into source (White) and three target (Black or African American, Asian, and Native American or Alaska Native) data sets.

We perform 10-fold cross validation to split the data into training and testing for the target data set of interest $T_{2}$.
The reported metrics are those calculated on the held out testing data while the training data are used to learn the RECaST posterior distribution or the baseline ridge parameter estimates.
We use 1 fold as the training and the remaining 9 as testing.
This process is repeated 10 times, until all of the target data have been used for training exactly once.
Note that the $T_{1}$ data is not split and all of it is used to learn the posterior distribution for $T_{1}$. 
The posterior predictive sampling values are set as $n_{\text{post}} = 50$, $n_{\bB} = 100$, and $n_{\bY} = 20$.
We perform three analyses corresponding to different target data sets to explore the relationship between racial groups.
This will help leverage the most similar data to build reliable target models.

\begin{table}[H]
    \centering
    \tiny
    \caption{HP Data feature and outcome summaries stratified by racial group. For age, CAL, and PD the 25$^{\text{th}}$, 50$^{\text{th}}$, and 75$^{\text{th}}$ percentiles are presented.}

    \begin{tabular}{c c c c c }
          & White & Black or African American & Asian & Native American or Alaska Native\\\hline
         $n$ & 21,029 & 1,297 & 1,050 & 153\\ \hline
         
         Age [25\%, 50\%, 75\%] & [46, 55, 67] & [37, 45, 54] & [36, 44.5, 57] & [41, 51, 59]\\
         Female & 12,577(59.8\%) & 744(57.4\%) & 617(58.8\%) & 102(66.7\%)
     \\ \hline
         
         Diabetes\\
         Not indicated & 19,460(92.5\%) & 1,148(88.5\%) & 959(91.3\%) & 134(87.6\%) \\
         Type 1  & 147(0.7\%) & 17(1.3\%)  & 7(0.7\%) & 2(1.3\%)\\
         Type 2 & 1,422(6.8\%) & 132(10.2\%) & 84(8\%) & 17(11.1\%) \\ \hline
         
         Tobacco Use \\ 
         Never & 16,909(80.4\%) & 1,062(81.9\%) & 946(90.1\%) & 114(74.5\%) \\
         Former & 2,200(10.5\%) & 89(6.9\%) & 37(3.5\%) & 21(13.7\%) \\ 
         Current & 1,920(9.1\%) & 146(11.3\%) & 67(6.4\%) & 18(11.8\%) \\ \hline
         
         Brushing \\
         Not indicated & 4,277(20.3\%) & 231(17.8\%) & 184(17.5\%) & 28(18.3\%) \\
         Daily & 16,635(79.1\%) & 1,057(81.5\%) & 863(82.2\%) & 124(81\%) \\
         Weekly & 96(0.5\%) & 8(0.6\%) & 3(0.3\%) & 1(0.7\%) \\ 
         Less than weekly & 21(0.1\%) & 1(0.1\%) & 0(0\%) & 0(0\%) \\ \hline
         
         Flossing \\
         Not indicated & 6,672(31.7\%) & 466(35.9\%) & 331(31.5\%) & 49(32\%) \\
         Daily & 7,321(34.8\%) & 404(31.1\%) & 434(41.3\%) & 50(32.7\%)\\
         Weekly & 4,583(21.8\%) & 229(17.7\%) & 174(16.6\%) & 39(25.5\%) \\ 
         Less than Weekly & 2,453(11.7\%) & 198(15.3\%) & 111(10.6\%) & 15(9.8\%) \\ \hline

         Insurance \\
         Commercial (C) &  16,186(77\%) & 1029(79.3\%)  & 806(76.8\%) & 125(81.7\%)\\
         Government Subsidized (G) &  2,594(12.3\%) & 232(17.9\%)  & 213(20.3\%) & 22(14.4\%)\\
         Private Pay (P)& 67(0.3\%) & 0(0\%) & 3(0.3\%) & 1(0.7\%)\\ 
         C + G & 312(1.5\%)  & 13(1\%)  & 1(0.1\%) & 1(0.7\%)\\
         C + P & 34(0.2\%)  & 1(0.1\%) & 0(0\%) & 0(0\%)\\
         G + P & 2,594(12.3\%)  & 232(17.9\%)  & 213(20.3\%) & 22(14.4\%)\\
         All & 10($<$0.1\%)  & 0(0\%) & 1(0.1\%) & 0(0\%)\\ \hline
         
         CAL [25\%, 50\%, 75\%] & [0.611, 1.518, 2.086] & [0.625, 1.766, 2.375] & [1, 1.821, 2.378] & [0.406, 1.413, 1.981]\\
          PD [25\%, 50\%, 75\%] & [1.552, 1.984, 2.302] & [2.032, 2.379, 2.745] & [1.763, 2.135, 2.458] & [1.625, 2.069, 2.431]\\ 
\\
         
    \end{tabular}
    \label{table: HP Data}
\end{table}

The first analysis uses the data from the Black or African American participants as $T_{1}$ and the Asian participants as the domain of interest $T_{2}$.
Based on the cross validation scheme, this gives a training size of 105 for each fold; there are many more training data points than features.
Table \ref{table: hp black asian} reports the predictive and coverage results.
We see that all of the multivariate transfer learning methods boast lower Mahalanobis distances when compared to the target only ridge baseline.
Univariate RECaST performs worse than both the baseline and the other transfer learning methods.
This again demonstrates the importance of considering the outcomes jointly.
All of the multivariate methods achieve desirable empirical coverage values slightly above the nominal 95\% level.
Univariate RECaST also achieves near nominal marginal coverage on the PD outcome, but an empirical coverage of 100\% on CAL indicating that at the 95\% nominal level the credible set contains the entire outcome space.

\begin{table}[h]
    \centering
    \caption{
    Performance metrics averaged over 10-fold cross validation.
    We report the joint empirical coverage at the 95\% nominal level; for univariate RECaST the marginal coverages are reported.
    For the online methods we report the posterior mean of $\alpha$ averaged across the fold. 
    The source population is the White participants. The target of interest is the Asian participants where 10\% of the data was used for training in each cross validation fold ($n_{T_{2}} = 105$).
    For the online methods, the informative prior is created using the corresponding multivariate RECaST method with the Black or African American participants.
    }

    \begin{tabular}{c|c c c}
        Model &  Mahalanobis Distance  & 95\% Coverage & $\mathbb{E}(\alpha  \mid \by_{T_{2}, 1}, \dots \by_{T_{2}, n_{T_{2}}})$ \\ \hline
        Target Ridge  & 1.22 (0.07)   \\
        Univariate & 1.54 (0.11) & 100, 97   \\
        MV Cauchy  & 1.21 (0.06)  & 98 \\
        MV-On Cauchy & 1.21 (0.07)  & 98 & 0.67 \\
        MV Copula& 1.22 (0.07)  & 97 \\
        MV-On Copula& 1.2 (0.06) & 97 & 0.33\\
    \end{tabular}
    \label{table: hp black asian}
\end{table}

Interestingly, the online methods have posterior means of $\alpha$ centered on opposite ends of the spectrum.
Since a uniform prior was used for $\alpha$, an average posterior mean of 0.67 corresponds to the online multivariate Cauchy method putting significant weight on the prior informed by the model built using the Black or African American participants.
An average posterior mean of 0.33 for the online copula-based method shows that it minimized the use of information from $T_{1}$.

The next analysis uses the data from the Black or African American participants as $T_{1}$ and the Native American or Alaska Native participants as the domain of interest $T_{2}$ with a training size of 15 for each fold.
Table \ref{table: hp black native} reports the predictive and coverage results where we see similar results as before.
All of the multivariate RECaST methods outperform the ridge regression with smaller Mahalanobis distances.
In this case, the MV-On Cauchy method has the best overall predictive performance.
Again, all of the multivariate methods achieve desired slightly conservative joint coverage.
Neither of the marginal coverages from univariate RECaST provide information at the 95\% level.

\begin{table}[H]
    \centering
    \caption{
    Performance metrics averaged over 10-fold cross validation.
    We report the joint empirical coverage at the 95\% nominal level; for univariate RECaST the marginal coverages are reported.
    For the online methods we report the posterior mean of $\alpha$ averaged across the fold. 
    The source population is the White participants.
    The target of interest is the Native American and Alaska Native participants where 10\% of the data was used for training in each cross validation fold ($n_{T_{2}} = 15$).
    For the online methods, the informative prior is created using the corresponding multivariate RECaST method with the Black or African American participants.
    }

    \begin{tabular}{c|c c c}
        Model &  Mahalanobis Distance  & 95\% Coverage & $\mathbb{E}(\alpha  \mid \by_{T_{2}, 1}, \dots \by_{T_{2}, n_{T_{2}}})$ \\ \hline
        Target Ridge  & 1.61 (0.64)  \\
        Univariate  & 1.92 (0.35) & 100, 100 \\
        MV Cauchy & 1.51 (0.25)  & 97 \\
        MV-On Cauchy & 1.46 (0.25)  & 98 & 0.67 \\
        MV Copula & 1.5 (0.26) & 97\\
        MV-On Copula & 1.5 (0.26) & 97 & 0.36  
        \end{tabular}
    \label{table: hp black native}
\end{table}

Finally, we use the data from the Asian participants as $T_{1}$ and the Native American or Alaskan Native participants as the domain of interest $T_{2}$.
Table \ref{table: hp asian native} reports the predictive and coverage results.
Again all multivariate RECaST methods provide better predictive performance than the baseline ridge. 
This data setting shows the biggest difference between the baseline and transfer learning methods with the MV-On Cauchy method again performing the best.
Empirical joint 95\% coverage is achieved by all of the multivariate RECaST methods; univariate RECaST again fails to provide useful uncertainty quantification for both CAL and PD.

\begin{table}[h]
    \centering
    \caption{
    Performance metrics averaged over 10-fold cross validation.
    We report the joint empirical coverage at the 95\% nominal level; for univariate RECaST the marginal coverages are reported.
    For the online methods we report the posterior mean of $\alpha$ averaged across the fold.
    The source population is the White participants. The target of interest is the Native American and Alaska Native participants where 10\% of the data was used for training in each cross validation fold ($n_{T_{2}} = 15$).
    For the online methods, the informative prior is created using the corresponding multivariate RECaST method with the Asian participants.
    }

    \begin{tabular}{c|c c c}
        Model &  Mahalanobis Distance & 95\% Coverage & $\mathbb{E}(\alpha  \mid \by_{T_{2}, 1}, \dots \by_{T_{2}, n_{T_{2}}})$  \\ \hline
        Target Ridge & 1.61 (0.64) \\
        Univariate  & 1.96 (0.79) & 100, 100\\
        MV Cauchy & 1.55 (0.58)  & 97 \\
        MV-On Cauchy & 1.47 (0.61)  & 97 & 0.67 \\
        MV Copula  & 1.5 (0.6)  & 97 \\
        MV-On Copula & 1.5 (0.6)  & 97 & 0.34 \\
    \end{tabular}
    \label{table: hp asian native}
\end{table}

To summarize, all of the multivariate RECaST methods improve joint prediction of CAL and PD in underrepresented populations.
The biggest improvements are seen on the Native American and Alaska Native population, which is the least represented population in the data.
The online approaches give the most accurate predictions, demonstrating the importance of including related target data.
All of the multivariate RECaST methods provided excellent joint coverage properties, which could provide clinicians both improved predictions and credible sets.

\section{Concluding Remarks}
\label{section: concluding remarks}

Future work for this method includes considering multiple measurements per participant.
Some dental data sets provide CAL and PD measurements for all of the teeth in a participant's mouth rather than a single averaged outcome \citep{guan2020bayesian}.
This could further expand into including modeling longitudinal data with multiple follow-up visits for each participant.
With the inclusion of other dental measurements, the spatial relationship between teeth can also be used \citep{jhuang2020spatiotemporal}.
Additionally, this multivariate framework can be applied to multiclass outcomes that are common for image classification \citep{deng2009}.

\section*{Acknowledgments}

This work was partially supported by the National Institutes of Health (R01-DE031134).
The authors also thank Dr. Dipankar Bandyopadhyay of Virginia Commonwealth University for access to the data and helpful discussions about the data analysis.
The last author is partially supported by the U.~S. National Science Foundation, grant DMS–2337943.

\appendix

\addcontentsline{toc}{section}{Appendices}
\renewcommand{\thesubsection}{\Alph{subsection}}

\section{Multivariate Cauchy Gibbs Sampler}
\label{section: gibbs}

For univariate RECaST, the following is the equivalent of putting a location-scale t-distribution prior on the mean of a Gaussian random variable
\begin{align*}
    \by_{1} , \dots , \by_{n_{T}} \mid \mu, \sigma^{2}, \sigma_{0}^{2} & \sim \mathcal{N}(\mu, \sigma^{2}) \\
    \mu \mid \sigma_{0}^{2} & \sim \mathcal{N}(\delta, \sigma^{2}_{0}) \\
    \sigma_{0}^{2} & \sim \text{IG}\Big(\frac{\nu}{2},\frac{\nu \cdot \gamma^{2}}{2}\Big). \\
    & \text{Then take priors} \\
    \delta & \sim \mathcal{N}(\mu_{\delta}, \sigma^{2}_{\delta}) \\
    \gamma & \sim \text{IG}(a_{\gamma}, b_{\gamma}).
\end{align*}

For the multivariate $t$-distribution, we have the following hierarchical representation:
\begin{align*}
    \bY_{T,1} , \dots , \bY_{T,n_{T}} \mid \bmu ,\bSigma & \sim \mathcal{N}_{m}(\bmu, \bSigma) \\
    \bmu \mid u, \bdelta, \bGamma & \sim \mathcal{N}_{m}(\bdelta , u^{-1} \bGamma) \\
   u & \sim \text{Gamma}\Big(\frac{\nu}{2}, \frac{\nu}{2}\Big) \\
   & \text{with priors} \\
   \bSigma & \sim \text{IW}_{m}(\bPsi_{\bSigma}, \nu_{\bSigma} ) \\
   \bdelta & \sim \mathcal{N}_{m}(\bmu_{\bdelta} , \bSigma_{\bdelta}) \\
   \bGamma & \sim \text{IW}_{m}(\bPsi_{\bGamma}, \nu_{\bGamma}),
\end{align*}

where $\nu$ is the degrees of freedom for the $t$-distribution prior on $\bmu$.
The following full conditional distributions can be used for Gibbs sampling
\begin{align*}
    %
    \bmu & \mid \bY_{T,1} , \dots , \bY_{T,n_{T}} , \bSigma , \bdelta, \bGamma, u \sim \mathcal{N}\Bigg(  \Big[ u \bGamma^{-1} + n_{T} \bSigma^{-1} \Big]^{-1} \Big[ u \bGamma^{-1} \bdelta + n_{T} \bSigma^{-1} \bar{\by_{T}} \Big], \Big[ u \bGamma^{-1} + n_{T} \bSigma^{-1} \Big]^{-1}\Bigg) \\
    %
    %
    \bSigma & \mid \bY_{T,1} , \dots , \bY_{T,n_{T}} , \bmu , \bdelta, \bGamma, u \sim \text{IW}\Bigg( n_{T} + \nu_{\bSigma} , \bPsi_{\bSigma} + \sum_{i=1}^{n_{T}} [\by_{T,i} - \bmu][\by_{T,i} - \bmu]^{\top} \Bigg) \\
    %
    %
    \bdelta & \mid \bY_{T,1} , \dots , \bY_{T,n_{T}} , \bmu , \bSigma, \bGamma, u \sim \mathcal{N}_{m}\Bigg( [\bSigma_{\bdelta}^{-1} + u \bGamma^{-1}]^{-1} [\bSigma_{\bdelta}^{-1} \bmu_{\bdelta} + u \bGamma^{-1} \bmu] , [\bSigma_{\bdelta}^{-1} + u \bGamma^{-1}]^{-1}\Bigg) \\
    %
    %
    %
    \bGamma & \mid \bY_{T,1} , \dots , \bY_{T,n_{T}} , \bmu , \bSigma, \bdelta, u \sim \text{IW}_{m}\Big( \nu_{\bGamma} - 1 ,  u(\bmu - \bdelta)(\bmu - \bdelta)^{\top} + \bPsi_{\bGamma} \Big) \\
    %
    %
    %
    u & \mid \bY_{T,1} , \dots , \bY_{T,n_{T}} , \bmu , \bSigma, \bdelta, \bGamma \sim \text{Gamma}\Big( \frac{\nu + m}{2}, \frac{1}{2} [\bmu - \bdelta]\bGamma^{-1}[\bmu - \bdelta] + \frac{\nu}{2} \Big).
\end{align*}

Setting $\nu = 1$ corresponds to setting a multivariate Cauchy prior on $\bmu$.

\section{Multivariate Gaussian Copula Finite Integrals}
\label{section: copula integrals}
 
\begin{align*}
    \pi & (\delta_{1}, \dots , \delta_{m} , \gamma_{1}, \dots , \gamma_{m}, \bR, \bSigma \mid y_{1,1}, \dots , y_{n_{T},m}, \widehat{\bTheta}_{S}) \\
    & = \int_{\mathbb{R}} \cdots \int_{\mathbb{R}} \pi (\delta_{1}, \dots , \delta_{m} , \gamma_{1}, \dots , \gamma_{m}, \bR, \bSigma , \beta_{1,1}, \dots ,\beta_{n_{T}, m}\mid y_{1,1}, \dots , y_{n_{T},m}, \widehat{\bTheta}_{S})
    \ d\beta_{1,1} \dots \ d\beta_{n_{T}, m} \\
   & \propto \pi(\delta_{1}, \dots , \delta_{m} , \gamma_{1}, \dots , \gamma_{m}, \bR, \bSigma) \cdot \\
   & \hspace{0.5cm}\prod_{i=1}^{n_{T}} \int_{0}^{1} \cdots \int_{0}^{1}  \pi  (y_{i,1}, \dots , y_{i,m} \mid F^{-1}_{\beta_{i,1}}(u_{i,1}), \dots ,F^{-1}_{\beta_{i,m}}(u_{i,m}), \bSigma  , \widehat{\bTheta}_{S}) \cdot \\ 
    & \hspace{1cm} c(u_{i,1}, \dots ,u_{i, m} \mid \delta_{1}, \dots , \delta_{m} , \gamma_{1}, \dots , \gamma_{m}, \bR)
    \ du_{i,1} \dots \ du_{i, m}.
\end{align*}

\section{Online Mean Derivation}
\label{section: online mean derivation}

When there are $\ell$ target data sets, the marginal posterior of $\balpha$ is 
\begin{align*}
p(\balpha \mid \by_{T_{\ell}, 1}, \dots \by_{T_{\ell}, n_{T_{\ell}}} ) & \propto \pi(\balpha) \cdot \Big[ 
            \alpha_{\ell} \cdot k_{\ell} + \sum_{i=1}^{\ell-1}\alpha_{i}\cdot  k_{i} \Big].
\end{align*}
Where $\pi(\balpha)$ is the prior on the vector $\balpha$, $\alpha_{\ell}$ is the weight associated with the uninformative prior term $k_{\ell}$ and $\alpha_{i}$ and $k_{i}$ are the weights associated and posterior distribution terms associated with $T_{i}$, respectively, for $i \in 1, \dots , \ell - 1$.
The normalizing constant is
\begin{align*}
    1 & = c \cdot \int_{\alpha_{1}} \dots \int_{\alpha_{\ell-1}}
        \pi(\balpha) \cdot \Big\{ 
            \Big(1-\sum_{i=1}^{\ell-1}\alpha_{i}\Big) k_{\ell} + \sum_{i=1}^{\ell-1}\alpha_{i} k_{i} \Big\} \ d\alpha_{\ell-1} \dots d\alpha_{1} \\ 
        & = c \cdot \Big\{
                k_{\ell} \int_{\alpha_{1}} \dots \int_{\alpha_{\ell-1}}
                \pi(\balpha) \ d\alpha_{\ell-1} \dots d\alpha_{1} +  (k_{i} - k_{\ell}) \int_{\alpha_{1}} \dots \int_{\alpha_{\ell-1}} 
         \pi(\balpha) \cdot \alpha_{i} \ d\alpha_{\ell-1} \dots d\alpha_{1}
           \Big\} \\ 
       & = c \cdot \Big\{
       k_{\ell} + \sum_{i=1}^{\ell-1}(k_{i} - k_{\ell}) \mathbb{E}_{\pi(\balpha)}(\alpha_{i})
       \Big\} \\ 
   c & = \Big\{
       k_{\ell} + \sum_{i=1}^{\ell-1}(k_{i} - k_{\ell}) \mathbb{E}_{\pi(\balpha)}(\alpha_{i})
       \Big\}^{-1}.
\end{align*}
The posterior mean of $\alpha_{j}$ is 
\begin{align*}
\mathbb{E}(\alpha_{j} \mid \by_{T_{\ell}, 1}, \dots \by_{T_{\ell}, n_{T_{\ell}}} ) & = 
  c \cdot \int_{\alpha_{1}} \dots \int_{\alpha_{\ell-1}} \alpha_{j} \cdot \pi(\balpha) \cdot \Big[ 
    \Big(1-\sum_{i=1}^{\ell-1}\alpha_{i}\Big) k_{\ell} + \sum_{i=1}^{\ell-1}\alpha_{i} k_{i} \Big] \ d\alpha_{\ell-1} \dots d\alpha_{1} \\
& = c \cdot \Big\{
  k_{\ell} \int_{\alpha_{1}} \dots \int_{\alpha_{\ell-1}} \alpha_{j} \cdot \pi(\balpha) \ d\alpha_{\ell-1} \dots d\alpha_{1} + \\
  & \hspace{1.5cm} (k_{j} - k_{\ell})\int_{\alpha_{1}} \dots \int_{\alpha_{\ell-1}} \alpha_{j}^{2} \cdot \pi(\balpha) \ d\alpha_{\ell-1} \dots d\alpha_{1} + \\
  & \hspace{1.5cm} \sum_{i=1, i\neq j}^{\ell-1} (k_{i} - k_{\ell})  \int_{\alpha_{1}} \dots \int_{\alpha_{\ell-1}} \alpha_{j} \cdot \alpha_{i} \cdot \pi(\balpha) \cdot \ d\alpha_{\ell-1} \dots d\alpha_{1}
  \Big\} \\
& = c \cdot \Big[
  k_{\ell} \cdot \mathbb{E}_{\pi(\balpha)}(\alpha_{j}) +
    (k_{j} - k_{\ell}) \Big\{\mathbb{E}_{\pi(\balpha)}(\alpha_{j})^{2} + \text{Var}_{\pi(\balpha)}(\alpha_{j}) \Big\} + \\
  & \hspace{1.5cm} \sum_{i=1,i\neq j}^{\ell-1}(k_{i}-k_{j}) \Big\{ \text{Cov}_{\pi(\balpha)}(\alpha_{i}, \alpha_{j}) + \mathbb{E}_{\pi(\balpha)}(\alpha_{i}) \mathbb{E}_{\pi(\balpha)}(\alpha_{j}) \Big\}
  \Big].
\end{align*}

\bibliographystyle{apalike}
\bibliography{references}

\end{document}